\documentclass[12pt,a4paper]{article}
\usepackage{amssymb}
\usepackage{amsmath}
\usepackage{amsfonts}
\usepackage{enumerate}
\usepackage{float}
\usepackage{natbib}
\usepackage{verbatim}
\usepackage{graphicx}
\usepackage{lscape}
\usepackage[autostyle]{csquotes}
\usepackage{bbm}
\usepackage[dvipsnames]{xcolor}
\usepackage{pdflscape,array,booktabs}
\usepackage[a4paper,bindingoffset=0.2in,left=1in,right=1in,top=1in,bottom=1in,footskip=.5in]{geometry}
\usepackage{caption}
\usepackage{multicol}
\usepackage{multirow}
\usepackage{booktabs}

\restylefloat{table}

\graphicspath{ {images/} }   

\DeclareMathAlphabet{\pazocal}{OMS}{zplm}{m}{n}

\begin{document}

\title{Analyzing Subjective Well-Being Data with Misclassification\thanks{%
We are most grateful to Nattavudh Powdthavee for various discussions and
suggestions especially at the early stage of this research. We thank Jo
Blanden, Valentina Corradi, Szabolcs De\'{a}k, Zoe Fannon, Tom\'{a}\v{s}
Jagelka, Anthoulla Phella, Luca Rondina, Yuya Sasaki, Matt Shum, Daniel
Wilhelm and the seminar participants at Academia Sinica and the RES Junior
Symposium 2019 for helpful comments and discussions. }}
\author{Ekaterina Oparina{\small {}}\thanks{\textit{{\small {}E-mail address}%
}{\small {}: e.oparina@surrey.ac.uk}} \\
University of Surrey \and Sorawoot Srisuma{\small {}}\thanks{\textit{{\small %
{}E-mail address}}{\small {}: s.srisuma@surrey.ac.uk}} \\
University of Surrey\\
}
\date{May 13, 2019 }
\maketitle

\begin{abstract}
We use novel nonparametric techniques to test for the presence of
non-classical measurement error in reported life satisfaction (LS) and study
the potential effects from ignoring it. Our dataset comes from Wave 3 of the
UK Understanding Society that is surveyed from 35,000 British households.
Our test finds evidence of measurement error in reported LS for the entire
dataset as well as for 26 out of 32 socioeconomic subgroups in the sample.
We estimate the joint distribution of reported and latent LS
nonparametrically in order to understand the mis-reporting behavior. We show
this distribution can then be used to estimate parametric models of latent
LS. We find measurement error bias is not severe enough to distort the main
drivers of LS. But there is an important difference that is policy relevant.
We find women tend to over-report their latent LS relative to men. This may
help explain the gender puzzle that questions why women are reportedly
happier than men despite being worse off on objective outcomes such as
income and employment.



\end{abstract}



\vspace{1.5in}

\setcounter{page}{1}\newpage

\section{Introduction}

Happiness or well-being economics first appeared in the economics literature
in the early 1970s, see \cite{van_praag_welfare_1971}, \cite%
{easterlin_does_1974}. This fast growing, yet sometimes polarizing, subject
studies causes and consequences of subjective well-being (SWB) and has
provided many interesting insights into what makes people happy. Some of
which have led to important policy lessons such as the idea that
unemployment in the Western society is largely involuntary (\cite%
{winkelmann_why_1998}), that reducing the rates of joblessness should take
priority over reducing the inflation rates (\cite{di_tella_preferences_2001}%
), that people partially adapt to serious disability over time (\cite%
{oswald_does_2008}), and that cigarette taxes actually improve the happiness
of the likely smokers (\cite{gruber_cigarette_2006}).

The central variable used in the well-being literature is life satisfaction
(LS). LS is originally designed to capture the respondent's global
well-being (\cite{diener_satisfaction_1985}). While LS has been shown to be
correlated with a range of economic factors such as health and unemployment
in expected ways, there is also ample evidence from the experimental
literature the reporting of LS is affected by confounding factors including
questionnaire designs, temporal factors such as mood or weather, and
pressures to provide socially desirable answers (see, e.g., \cite%
{schwarz_mood_1983}, \cite{feddersen_subjective_2016}, \cite%
{schwarz_cognition_2014}). We can therefore view \textit{reported} LS as a
possible mismeasurement of \textit{latent} LS. Given the discrete nature of
the SWB responses measurement error is also known as a misclassification.

Misclassification is a form of non-classical measurement error. A
mismeasured LS can cause bias in empirical studies in arbitrary way. Indeed,
one particular takeaway from the well-known article by \cite%
{bertrand_people_2001}, entitled: \textquotedblleft Do people mean what they
say? Implications for Subjective Survey Data\textquotedblright , suggests
researchers should not use LS as a dependent variable. Nevertheless,
understanding the determinant of LS is one of the most fundamental tasks in
the well-being literature. Since there is no obvious solution to the
measurement error problem, LS is still routinely used as the dependent
variable and the potential effects from measurement error have been
unaccounted for.

In this paper we use novel econometric techniques to formally test for the
presence of measurement error in reported LS and, if it exists, account for
it and study its potential effects. We use survey data of 35,000 British
households from the UK\ Understanding Society taken between January 2011 and
June 2013. This (Wave 3) dataset is unique in that it contains what we
believe are suitable variables that enable us to test for measurement errors
and use the misclassification model of \citet{hu_identification_2008} to
identify the joint distribution of the reported and latent LS
nonparametrically. In particular, the LS distribution will be able to
provide insights into the (mis-)reporting probabilities for people of
different demographic and socioeconomic groups. We can also use this
distribution to identify the determinants of latent LS in popular parametric
models in the literature such as linear projection and ordered response
models (e.g. logit and probit) without observing latent LS. Our results can
have important policy implications, whether it is for the purpose of helping
policy makers identify suitable groups of individuals for an intervention or
for quantifying impacts of policies based on latent LS as opposed to
reported LS.

We emphasize at this point that the empirical results in our paper are free
from recent criticisms on the general econometric analysis of SWB data. In
particular, \citet{bond_sad_nodate} show the mean ranking of happiness or LS
generally cannot be identified unless strong assumptions (such as
homoskedasticity in probit/logit models) are a priori imposed. An
implication of this is the sign of a parameter estimate is not informative
on the average effect. Subsequently, \citet{bond_sad_nodate} use an
(heteroskedastic) ordered probit to show some of the most well-known results
in the happiness literature can be arbitrarily reversed. Chen, Oparina,
Powdthavee and Srisuma (\citeyear{chen_have_2019}) point out a simple
solution that is to focus on the median effect instead of the mean. In this
paper we use a heteroskedastic ordered probit model without specifying the
form of heteroskedasticity parametrically. Our estimates and partial effects
are to be interpreted through the median accordingly.

We begin our empirical study by testing for the presence of measurement
errors in reported LS. We adopt the nonparametric approach suggested
recently by \cite{wilhelm_testing_2018} that, under suitable conditions,
stochastic dependency between some auxiliary variables conditioning on the
reported variable can signify the presence of measurement errors.\footnote{%
The insight that conditional independence can indicate the presence of
measurement error was first explored in a regression context by \cite%
{mahajan_identification_2006}, who considers a binary regressor that may be
measured with error.} We use a Kolmolgorov-Smirnov type statistic and find
evidence of measurement error in the reported LS for the entire dataset as
well as for 26 out of 32 socioeconomic subgroups in the sample.

Next, we estimate a model of latent life satisfaction conditioning on key
socioeconomic variables. We find the main drivers of LS in the model with
latent LS are the same as those with reported LS, suggesting that the bias
from measurement error may not be substantial enough to distort the effects
of the main factors. For example, marriage and health have clear positive
impact on LS while income and education have insignificant effects that
would otherwise be positive due to substitution effects with health.
However, there is one notable difference. We find that women systematically
report themselves to be more satisfied with lives than they actually are
relative to men. Measurement errors may thus help us solve the \textit{%
gender puzzle} that women are happier than men in spite of the fact that
they are often associated with less favorable objective measures in terms of
health, income and employment level (see \cite{dolan_we_2008}, \cite%
{stevenson_paradox_2009}).

The validity of our empirical results relies on the conditions of the
misclassification model of \citet{hu_identification_2008} being satisfied.
Hu's identification procedure requires assumptions on the respondents'
reporting behavior as well as two auxiliary variables. These variables need
to be independent to each other and to the reported LS conditional on the
latent LS.\footnote{%
The same conditional independence assumption between the auxiliary variables
is also required for the measurement error test.} The auxiliary variables
also need to be appropriately correlated with the latent variable. These
requirements pose contrasting qualities for candidates for auxiliary
variables, which is an empirical challenge somewhat analogous to finding a
good instrument. The selection of appropriate auxiliary variables requires a
transparent interpretation of the origin of misreporting.

In this paper, we focus on external factors such as the effects of
questionnaire design, temporal factors, and socially-desirable responding as
the source of measurement errors. These factors influence the reporting
behavior, yet at the same time are idiosyncratic in the sense that they do
not contribute to the LS in general. Subsequently, the two auxiliary
variables we select for identification are: (i) a measure of mental state
that is derived from General Health Questionnaire (GHQ); (ii) a derived
measure of neuroticism, which is one of the traits that underlies one's
personality. The latter in particular is currently collected only in Wave 3
of the UK Understanding Society survey. We provide a detailed discussion of
the conditions for identification and the logic behind our choice of
auxiliary variables in the paper. Ultimately, similarly to the exogeneity of
an instrument, the conditional independence of the variables is an
untestable identifying assumption. A nice feature of our dataset is that we
have information for all questions that are used to derive measures from GHQ
and neuroticism. Some of these questions are more objective than others,
which allow us to perform robustness checks by constructing different
versions of auxiliary variables that intuitively have a varying degree of
trade-off between independence and relevance.

We make three main contributions in this paper. (1) To the best of our
knowledge, we are the first to apply the nonparametric test of \cite%
{wilhelm_testing_2018} and misclassification model of %
\citet{hu_identification_2008} to analyze SWB data. These novel econometric
methods do not rely on unjustified parametric assumptions, allow for
non-classical errors, and do not require validation data (cf. \cite%
{bound_extent_1991}, \cite{chen_measurement_2005}). The latter two features
in particular seem to be necessary for making any progress in accounting for
measurement errors in self-reported subjective variables. (2) Our empirical
study shows statistically that measurement errors exist in reported LS,
using Wave 3 of the UK\ Understanding Society data. While we find some
similarities between the relations between reported and latent LS with key
covariates there are important differences. Measurement errors can therefore
have practical implications if policy makers are to make decisions based on
reported as opposed to latent LS. (3) We bridge the gap between theory and
practice. Modern econometrics studies on nonparametric identification of
models with measurement errors tend to be mathematically sophisticated and
they aim to identify the joint distribution of all variables in the model.
Most empirical researchers, especially those currently dealing with data
that may be measured with errors, on the other hand employ parametric
models. We show how nonparametric identification results can lead to
parametric identification in popular models that are used to analyze SWB
data in the literature.

We organize the rest of the paper as follows. Section 2 gives a brief
background on the current use of SWB data. Section 3 presents the
econometric model, gives conditions for identification of the parameters of
interest, and discuss practical inference. Section 4 describes the test we
use to detect possible measurement errors in the reported LS. The empirical
application is in Section 5. Section 6 concludes. The Appendix provides more
detailed description of the data and supplementary results to support the
findings in Section 5.

\section{Background}

Our background section consists of three parts. Section 2.1 provides a brief
overview for a measure of well-being. Section 2.2 summarizes the main
approaches to analyzing SWB data as well as some recent criticisms. Section
2.3 discusses measurement errors in self-reported well-being variables.

\subsection{Subjective well-being}

Well-being research is motivated by the ambition to understand the key
drivers of individual's well-being. SWB is an umbrella term that includes a
person's \textit{cognitive well-being} such as LS (i.e., a judgment one
makes about one's life overall), \textit{affective well-being} (i.e.,
frequency and intensity of experienced emotions), and \textit{eudaimonic
well-being} (i.e., sense of purpose and worthwhileness). The economics of
happiness literature traditionally uses LS, rather than measures of
emotional states, as a proxy utility data.

Here we list some of the stylized facts of this literature as summarized in
the World Happiness Report (\cite{helliwell_world_2012}):

\begin{itemize}
\item Richer people are on average happier than poorer people;

\item LS is highly positively correlated with mental and physical health;

\item Marriage has a positive correlation with LS;

\item LS is U-shaped in age;

\item Unemployment is significantly detrimental to LS;

\item In most developed countries women report higher LS than men, despite
being worse off in measurable socioeconomic outcomes;

\item There is little correlation between a person's education level and
his/her LS, but education is indirectly related to happiness through its
effect on income: education increases income and income increases happiness.
\end{itemize}

Given the subjective nature of LS, the overwhelming majority of the findings
is based on self-reported assessment: respondents are asked to report how
satisfied they are with their life on a given scale. This approach favors
personal evaluation of global well-being over the views of potential
experts. Despite earlier concerns, self-reported measures of life
satisfaction are proven to have a degree of validity. They converge in
expected ways with each other and with non-self-reported measures, such as
those based on other people's reports and the behavior of the respondent (%
\cite{diener_assessing_2009}). They are also predictive of future behaviors,
such as job quit, divorce, and suicide (\cite{diener_findings_2017}).

\subsection{Estimating life satisfaction}

The most common feature of empirical studies in the well-being literature is
to use reported LS as a dependent variable and other characteristics, such
as income, gender, health and employment statuses etc., as covariates. There
are two distinct approaches in how life satisfaction is modeled. One treats
LS as a \textit{cardinal} variable and the other as an \textit{ordinal} one.
The statistical techniques used for the former are based on least squares
estimation or direct comparisons between sample averages. For the ordinal
case, ordered logit or probit models are typically used. Both approaches are
widely used in practice. See \cite{ferrericarbonell_how_2004} for an account
for some (dis-)similarities of results between the two approaches.

The econometric analysis of SWB data has come under recent heavy criticisms.
Whether least squares regression or ordered probit/logit estimation is used,
similar to most other economic fields, a typical approach researchers take
is to then draw conclusions based on statements about the relative \textit{%
mean} happiness between groups of individuals (e.g. men and women, employed
and unemployed, or across countries etc). Critiques point out that these
research ignore the fact that SWB data are ordinal in nature. And the mean
ranking of ordinal variables is only identified when it is stable across all
increasing transformation. For examples, this means unless relevant
stochastic dominance conditions hold, the raw average ranking and signs of
least squares estimates may be reversed by monotonically transforming the
ordinal scale, see \citet{schroder_revisiting_2017} and %
\citet{bond_sad_nodate}. Importantly, this issue goes deeper than
\textquotedblleft using OLS to estimate a discrete dependent
variable\textquotedblright , as \citet{bond_sad_nodate} also show the mean
ranking of latent happiness from ordinal models can also be arbitrarily
reversed. They use a heteroskedastic ordered probit model to illustrate it
for some of the most well-known results in the happiness literature.
Explicitly allowing for heteroskedasticity is important because a
homoskedastic model a priori effectively assumes the mean ranking to be
identified. See Theorem 1 of \cite{bond_sad_2014}.

There are ways to analyze happiness data that avoid these criticisms. For
examples, direct comparisons of probabilities or probability odds of certain
events between groups are not affected (e.g., \cite{easterlin_will_1995}).
But such a descriptive approach has limited scope for incorporating
covariates. Chen, Oparina, Powdthavee and Srisuma (\citeyear{chen_have_2019}%
) suggest one solution is to use the \textit{median} instead of the \textit{%
mean} as a mode of comparison. The median rank is stable across all
increasing transformations. Furthermore, they highlight the fact that the
median and the mean in symmetric parametric models, like probit and logit,
are the same. The median has therefore been frequently estimated but only
interpreted as the mean.\footnote{%
Estimating the median without any parametric distributional assumption is
also possible (\cite{manski_semiparametric_1985}, \cite{lee_median_1992}). 
\cite{chen_have_2019} suggest the semiparametric median can be estimate
using modern constrained mixed integer optimization technique; they apply it
to study the Easterlin paradox.} This fact instantly nullifies the reversal
of prior results in \citet{bond_sad_nodate}\ by simply interpreting those
estimates through the median. To this end, our paper emphasizes the use of
an ordered response model with heteroskedasticity. We show in Section 4 it
is in fact simple to estimate a heteroskedastic probit model even without
specifying the form of the heteroskedasticity parametrically.

\subsection{Measurement error}

It is the norm in practice to assume that LS is measured without errors. In
this work we take the view that \textit{reported} LS is a combination of 
\textit{latent} LS and measurement error:\footnote{%
In this paper we use the term \textit{latent} to mean a measurement without
error. In Section 3.2.2 we model $X^{\ast }$\ using an ordered response
model, which traditional interprets $X^{\ast }$ to be derived from an
underlying continuous happiness variable that is plays an analous role to
utility in McFadden's random utility maximization model.} 
\begin{equation*}
X=X^{\ast }+u.
\end{equation*}%
We denote the measurement error by $u$. Since $X$ and $X^{\ast }$\ are
discrete, $u$ is also discrete. This type of measurement error is also known
as misclassification. Misclassification is non-classical by nature. For
example, given the number of values the variable can take is finite, extreme
values can only be mismeasured in one direction so a zero-mean error
(conditional on the true value) is impossible. Furthermore, the error term
is likely to be correlated with the covariates that are typically used in LS
analysis.

The error in LS is thought to come from two main distinct sources. One is
the effects that influence the respondent's judgment about the level of LS
while it is being formed, such as passing effects and the effects of survey
design. The other comes from factors that influence how the respondents
communicate their judgment, i.e. social desirability bias. We now describe
these two sources in details.

Well-being research is typically interested in the relatively stable
well-being level, rather than in the passing effects, which is why
influences of temporal factors can be considered as measurement error that
should be controlled for. LS is theorized to be a judgment that a respondent
constructs while answering the question, so it can be influenced by temporal
factors that take place at the time of forming the judgement (\cite%
{strack_subjective_1991}). Multiple experiments have shown that this measure
can be influenced by mood manipulations like finding a dime in a copy
machine, receiving a chocolate bar, spending time in a pleasant environment
or watching a football team win (\cite{strack_subjective_1991}). In a
large-scale survey setting, mood swings can be caused by weather at the time
of the well-being judgment; there are well-known diurnal and day-of-the-week
variations in SWB (see \cite{diener_advances_2018}).

Research on survey design shows that respondents' answers can be manipulated
to some extent, see \cite{bertrand_people_2001}. Respondents tend to provide
answers consistent with the previous ones, so the ordering of questions
matters. Changing the wording of the question also affects the way people
respond to them, particularly when people are asked to agree or disagree
with a statement. When respondents are asked to assess a statement within a
given scale, answers can differ if different scales are provided. Error of
this type induces systematic bias, however it can be minimized by the
appropriate design on the questionnaire (\cite{oecd_oecd_2013}). Nowadays,
the majority of surveys have been designed taking this issues into account.

There are also concerns about respondents not reporting truthfully in an
endogenous way, i.e. error is correlated with regressors. Specifically,
respondents may modify their answers to make them seem more socially
desirable. E.g. \cite{diener_response_1991} refer to one version of social
desirability, specific to reported SWB, to be 
\enquote{happy image
management} that would result in reporting higher or lower well-being than
experienced to appear happier/less happy. Measurement errors of this kind
would make it difficult for us to distinguish between the case when
unemployed people are truly unsatisfied with their life or when they report
a lower score because being unemployed is associated with a less desirable
social status.

Measurement errors from both sources might be correlated with regressors,
i.e. respondents from different groups might systematically exhibit
different reporting behaviors. \cite{barrington-leigh_impact_2017} use two
major health surveys in Canada to show that women and individuals with poor
health condition are more affected by weather conditions. \cite%
{heffetz_conclusions_2013} use the reported number of call attempts made to
participants in the University of Michigan's Surveys of Consumers to show
the difference in reported happiness among easy-to-reach and hard-to-reach
respondents.

The discussion above suggests that statistical analysis using life
satisfaction is likely to be biased in some unknown ways if measurement
error is ignored. Some examples of these are highlighted in \cite%
{bertrand_people_2001}. Over the past fifteen years, the econometrics
literature has made advances on identifying nonclassical measurement error
without additional measurement or validation data on the mismeasured
variable. The approach we take in this paper follows from the
misclassification model of \cite{hu_identification_2008} that assumes all
the variables in the model are discrete. The discrete setup is suitable for
analyzing LS as most variables that are used in this literature are discrete
or can naturally be discretized. A more general treatment that allows for
some continuous variables can be found in \cite{hu_instrumental_2008}. We
refer the reader to the surveys by \cite{schennach_2013} and \cite%
{hu_econometrics_2017} for examples of applications that rely on this type
of identification results.

\section{Model and identification strategy}

In this section we describe a model of misclassification \cite%
{hu_identification_2008} in the context of our application. In Section 3.1
we introduce the variables, provide and discuss the assumptions required on
them, and outline the nonparametric identification strategy. We consider
parametric identification in Section 3.2. Section 3.3 discusses the
numerical aspects of estimation and inference.

\subsection{Nonparametric identification}

\label{Identification}Let $X^{\ast }$\ denote the latent LS. Suppose $%
X^{\ast }$ can take the following values: 
\begin{equation*}
X^{\ast }=%
\begin{cases}
1\quad \text{dissatisfied} \\ 
2\quad \text{neither satisfied nor dissatisfied} \\ 
3\quad \text{satisfied}%
\end{cases}%
.
\end{equation*}%
We assume to have three observed variables $\left( X,Y,Z\right) $. $X$ is
the reported LS. $Y$ is a derived measure of neuroticism that is indicative
of a responder's emotional stability.\footnote{%
Neurotic individuals can be defined by such terms as worrying, insecure,
self-conscious, and temperamental (\cite{mccrae_validation_1987}).} $Z$ is a
measure of mental states that is derived from General Health Questionnaire
(GHQ-12). We assume that $X$ and $X^{\ast }$ have the same support. $Y$ is a
binary indicator that takes a value of $1$ for the individuals whose level
of neuroticism is above the median of the sample and $0$ otherwise. The
support of $Z$ has the same cardinality as the support of $X$, which in this
case is $\left\{ 1,2,3\right\} $, ranging from $1$ -- not distressed to $3$
-- distressed.

We provide a particular description of the variables above to fix ideas,
which will be useful for motivating abstract assumptions of the
misclassification model. All of our assumptions and theoretical results
below are written in a more general term. More specifically, they are all
valid for $X^{\ast }$\ that takes values from any finite set as long as the
cardinality of the support of $\left( X^{\ast },X,Z\right) $ are the same.%
\footnote{%
If the cardinalities of the support of $X$ and $Z$\ are unequal initially,
one can always coarsen the data to satisfy the same cardinality condition.}$%
^{,}$\footnote{%
The setup that $Y$\ takes only two values is a minimal assumption for
identification. We can always convert any random variable into a binary
variable.}

In what follows, we will use $f_{A|B}\left( a|b\right) $ to denote $\Pr %
\left[ A=a|B=b\right] $ for random vectors $A$ and $B$ taking values $a$ and 
$b$ respectively, and $f_{A}\left( a\right) $ to denote the $\Pr \left[ A=a%
\right] $. We will denote a generic matrix whose $ij-$th element is $m_{ij}$%
\ by a bold font $\mathbf{M}:=\left( m_{ij}\right) $ and a diagonal matrix
with the $i-$th diagonal element $d_{i}$ by $\mathbf{D}:=diag\left\{ \left(
d_{i}\right) \right\} $. We denote a transpose of matrix $\mathbf{M}$ by $%
\mathbf{M}^{\top }$ and\ an inverse of an invertible matrix $\mathbf{M}$ by $%
\mathbf{M}^{-1}$. Correspondingly, when $A$ and $B$ are scalar variables
supported on $\left\{ a_{1},\ldots ,a_{d_{A}}\right\} $\ and $\left\{
b_{1},\ldots ,b_{d_{B}}\right\} $\ respectively, we then define $\mathbf{M}%
_{A|B}$ to be a $d_{A}$ by $d_{B}$ matrix such that $\mathbf{M}%
_{A|B}:=\left( f_{A|B}\left( a_{i}|b_{j}\right) \right) $; we define $%
\mathbf{M}_{A,B}$ similarly so that $\mathbf{M}_{A,B}:=\left( f_{A,B}\left(
a_{i},b_{j}\right) \right) $.


We assume $\left( X^{\ast },X,Y,Z\right) $ satisfies the following
conditions:

\bigskip

\textbf{Assumption 1 (CI).} $\left( X,Y,Z\right) $\textit{\ are independent
conditional on }$X^{\ast }$\textit{, i.e. } 
\begin{equation*}
X\enskip\bot \enskip Y\enskip\bot \enskip Z\enskip|\enskip X^{\ast }.
\end{equation*}


\textbf{Assumption 2 (RNK).} $\mathbf{M}_{X,Z}:=\left(f_{X,Z}%
\left(x_{i},z_{j}\right)\right)$\textit{\ has full rank.}

\bigskip{}

\textbf{Assumption 3 (UNQ).} $E[Y|X^{\ast }=x_{i}^{\ast }]$\textit{\ is
different for different }$i$\textit{.}

\bigskip{}

\textbf{Assumption 4 (ORD).} $f_{X|X^{\ast }}(x_{I}|x_{i}^{\ast })$ \textit{%
is strictly increasing in }$i=1,\ldots ,I$

\bigskip{}

Assumption 1 is the key \textit{conditional independence} assumption. While
it is easy to find three independent variables in isolation, the challenge
is to also have them satisfy Assumptions 2 to 4. We first explain why our
choice of $\left( X,Y,Z\right) $ may reasonably satisfy Assumption 1.
Suppose the source of the misclassification error that makes $X$ different
to $X^{\ast }$ comes from temporal factors (e.g. mood or weather), socially
desirable responding (e.g. {happy image management}) or questionnaire
design. We have selected $Z$ and $Y$ carefully so they contain information
on life satisfaction and some other information that we treat as errors. We
want these errors to be independent from the errors in $X$ and between
themselves once we control for $X^{\ast }$. For $Y$, the measure of
neuroticism is also constructed from the answers to multiple questions.
These questions concern personal traits rather than direct assessment of LS.
The answers are unlikely to be influenced, for example, by {happy image
management}, because the reporting of LS is different from experiences
regarding emotional stability. The questions about personal traits and those
about LS are also often asked in different parts of the survey (as is the
case for our dataset). It is designed to minimize the influence of questions
and answers that LS and neuroticism may have on each other. Moreover,
temporary factors such as weather, are less likely to influence individuals
evaluation of how often one worries, compared to the judgment about to what
extent she is satisfied with one's life. For $Z$, unlike the LS question,
the GHQ-12 measure is constructed from multiple questions with a varying
degree of subjectivity. Parts of the questions are as subjective as the
well-being question, e.g. 
\enquote{Have you recently been feeling
reasonably happy}, however, some questions ask for objective information
such as the amount of sleep. Given the questions are less subjective, the
answers are less likely to be influenced by similar cognitive effects. We
perform a robustness check on our estimation results by using different
GHQ-12 measures, which vary in degree of objectivity/subjectivity, in
Appendix B.

Assumption 2, unlike the other assumptions, is testable as it is a condition
on the observable. $\mathbf{M}_{X,Z}$ is a square matrix since $X$ and $Z$
have the same number of support points. The full rank condition is the
discrete analog to the \textit{completeness} assumption (see 
\citet[Assumption
3]{hu_instrumental_2008}), which ensures invertibility of $\mathbf{M}_{X,Z}$.

Assumptions 3 and 4 have more intuitive interpretations. Assumption 3 says
that the probability that an individual whose level of neuroticism is above
the median of the sample differs across sub-populations partitioned by $%
X^{\ast }$. Since neuroticism captures personal trait, which has been shown
to be strongly related to the level of LS (see, e.g., \cite%
{diener_subjective_2009}), this condition is likely to hold. Assumption 4
imposes monotone likelihood towards positive reporting. In particular, if we
set $x_{i}=I$ and $x_{i}^{\ast }=i$ for all $i$, respondents who are
latently {satisfied} with their lives are more likely to report the higher
state that those who are {neither satisfied nor dissatisfied}; analogously,
those who are latently {neither satisfied nor dissatisfied} are more likely
to report the higher state than those who are {not satisfied}.

\bigskip

In order to provide further insights on why A1 - A4 enable the
identification of $f_{X^{\ast },X,Y,Z}$, we now provide an intuitive outline
for the proof of Theorem 1 in \cite{hu_econometrics_2017}.

\subsubsection*{Identification of $f_{X^{\ast },X,Y,Z}$ under A1 - A4}

Under Assumption 1, we have:%
\begin{equation*}
f_{X^{\ast },X,Y,Z}=f_{X|X^{\ast }}f_{Y|X^{\ast }}f_{Z|X^{\ast }}f_{X^{\ast
}}.
\end{equation*}%
The distribution of $\left( X^{\ast },X,Y,Z\right) $\ is identified if we
can identify the distribution of $X^{\ast }$ and the marginal distributions
of $X,Y$ and $Z$\ conditional on $X^{\ast }$. Using the Law of Total
Probability, under Assumption 1, we have 
\begin{equation*}
f_{X,Y,Z}(x,y,z)=\sum_{x^{\ast }\in \pazocal{X}^{\ast }}f_{X|X^{\ast
}}(x|x^{\ast })f_{Y|X^{\ast }}(y|x^{\ast })f_{Z|X^{\ast }}(z|x^{\ast
})f_{X^{\ast }}(x^{\ast }),
\end{equation*}%
where we denote the support of $X^{\ast }$ by $\pazocal{X}^{\ast }$. Fix $%
Y=y $, we can define $\mathbf{M}_{X,y,Z}:=\left(
f_{X,Y,Z}(x_{i},y,z_{j})\right) $\ so that the above relation can be
vectorized for each $y$, 
\begin{equation}
\mathbf{M}_{X,y,Z}=\mathbf{M}_{X|X^{\ast }}\mathbf{D}_{y|X^{\ast }}\mathbf{D}%
_{X^{\ast }}\mathbf{M}_{Z|X^{\ast }}^{\top },  \label{A1}
\end{equation}%
where $\mathbf{M}_{X|X^{\ast }}:=\left( f_{X|X^{\ast }}(x_{i}|x_{j}^{\ast
})\right) ,\mathbf{D}_{y|X^{\ast }}:=diag\left\{ \left( f_{Y|X^{\ast
}}(y|x_{i}^{\ast })\right) \right\} ,\mathbf{D}_{X^{\ast }}:=diag\left\{
\left( f_{X^{\ast }}(x_{i}^{\ast })\right) \right\} $ and $\mathbf{M}%
_{Z|X^{\ast }}:=\left( f_{Z|X^{\ast }}(z_{i}|x_{j}^{\ast })\right) $.
Similarly, using the Law of Total Probability and Assumption 1, we can also
write 
\begin{equation*}
f_{X,Z}(x,z)=\sum_{x^{\ast }\in \pazocal{X}^{\ast }}f_{X|X^{\ast
}}(x|x^{\ast })f_{Z|X^{\ast }}(z|x^{\ast })f_{X^{\ast }}(x^{\ast }),
\end{equation*}%
which can be represented in a matrix notation by 
\begin{equation}
\mathbf{M}_{X,Z}=\mathbf{M}_{X|X^{\ast }}\mathbf{D}_{X^{\ast }}\mathbf{M}%
_{Z|X^{\ast }}^{\top },  \label{A2}
\end{equation}%
for $\mathbf{M}_{X,Z}:=\left( f_{X,Z}(x_{i}|z_{j})\right) $. When Assumption
2 holds, we have 
\begin{equation*}
\mathbf{M}_{X|X^{\ast }}^{-1}\mathbf{M}_{X,Z}=\mathbf{D}_{X^{\ast }}\mathbf{M%
}_{Z|X^{\ast }}^{T}.
\end{equation*}%
The above display can be used to combine (\ref{A1}) and (\ref{A2}) and
obtain $\mathbf{M}_{X,y,Z}=\mathbf{M}_{X|X^{\ast }}\mathbf{D}_{y|X^{\ast }}%
\mathbf{M}_{X|X^{\ast }}^{-1}\mathbf{M}_{X,Z}$, so that 
\begin{equation}
\mathbf{M}_{X,y,Z}\mathbf{M}_{X,Z}^{-1}=\mathbf{M}_{X|X^{\ast }}\mathbf{D}%
_{y|X^{\ast }}\mathbf{M}_{X|X^{\ast }}^{-1}.  \label{A3}
\end{equation}%
Hu's main insight is $\mathbf{M}_{X,y,Z}\mathbf{M}_{X,Z}^{-1}$, which is
identified by the data, can identify $\mathbf{M}_{X|X^{\ast }}\mathbf{D}%
_{y|X^{\ast }}\mathbf{M}_{X|X^{\ast }}^{-1}$ by an eigen-decomposition,
where the diagonal elements of $\mathbf{D}_{y|X^{\ast }}$ are the
eigenvalues and $\mathbf{M}_{X|X^{\ast }}$\ is a matrix of the corresponding
eigenvectors. Assumptions 3 and 4 ensure that the eigen-decomposition
produces a unique and distinct ordering of eigenvalues, thus $f_{X|X^{\ast
}} $ and $f_{Y|X^{\ast }}$ are identified. In turn they also identify $%
f_{Z|X^{\ast }}$ and $f_{X^{\ast }}$. To see this, first note that $%
f_{X}(x)=\sum_{x^{\ast }\in \pazocal{X}^{\ast }}f_{X|X^{\ast }}(x|x^{\ast
})f_{X^{\ast }}(x^{\ast })$, so that we can identify $f_{X^{\ast }}$\ by
(pre-)multiplying a vector of $\left( f_{X}\left( x_{i}\right) \right) $\ by 
$\mathbf{M}_{X|X^{\ast }}^{-1}$. Then $f_{Z|X^{\ast }}$\ can be identified
by solving, for instance, equation (\ref{A1}) for $\mathbf{M}_{Z|X^{\ast
}}^{T}$. Thus $f_{X^{\ast },X,Y,Z}$\ is identified when Assumptions 1 to 4
hold.

\bigskip

The argument above makes clear that we use Assumptions 3 and 4 only for the
purpose of identifying the eigen-decomposition of $\mathbf{M}_{X,y,Z}\mathbf{%
M}_{X,Z}^{-1}$. While we cannot test Assumptions 3 and 4 directly, in
practice we can estimate $\mathbf{D}_{y|X^{\ast }}$\ and $\mathbf{M}%
_{X|X^{\ast }}$ without fully imposing Assumptions 3 and 4 a priori. For
example, if the inequalities in Assumption 4 are violated empirically then
this would suggest some conflicts with the data. (We provide more discussion
on this in Section 3.3 and Section 5.) In this case one should seek other
economically plausible conditions to ensure uniqueness of the
eigen-decomposition. Alternative conditions for identification can be found
in \cite{hu_identification_2008}.

The identification of $f_{X^{\ast },X,Y,Z}$\ \ gives a complete
characterization of the stochastic relation between all the variables in the
model. In particular, the model offers new insights into various reporting
behaviors conditioning on the latent level of satisfaction. Next we show how 
$f_{X^{\ast },X,Y,Z}$ can be used in conjunction with additional covariates
to identify commonly used parametric models.

\subsection{Parametric identification}

Empirical studies are most often interested in the coefficients in linear
and probit/logit models of LS given a vector of covariates\ $Q$. If we use
only reported LS then these parameters can be written as some functionals of 
$f_{X,Q}$, which are identified from the observed data under familiar
conditions. We would like to identify and estimate analogous parameters for
latent LS. This is possible even if we do not observe latent LS as long as $%
f_{X^{\ast },Q}$\ is identified.

Since $Q$ may not necessarily contain $\left( Y,Z\right) $, which are used
for nonparametric identification, we write $Q=\left( R^{\top },W^{\top
}\right) ^{\top }$\ so that $R$, if it is non-empty, contains either $Y$
and/or $Z$, and $W$\ is a vector of all other conditioning variables. We
shall assume throughout that $W$\ is a discrete random variable. So that all
variables in the model are discrete and take values from some finite set. We
assume our data satisfy the following condition.

\bigskip

\textbf{Assumption A.} $\left\{ \left( X_{n},Y_{n},Z_{n},W_{n}\right)
\right\} _{n=1}^{N}$ \textit{is a random sample of }$\left( X,Y,Z,W\right) $%
\textit{\ with }$N\rightarrow \infty $\textit{\ such that }$\left(
X,Y,Z\right) $\textit{\ conditional on }$W$\textit{\ satisfies Assumptions 1
to 4 almost surely.}

\bigskip

Assumption A ensures that $f_{X^{\ast },X,Y,Z|W}$\ is identified. This
allows us to identify $f_{X^{\ast },Q}$.

\bigskip

\textbf{Lemma 1. }\textit{Suppose Assumption A holds. Then }$f_{X^{\ast },Q}$%
\textit{\ is identified.}\ 

\textbf{Proof. }The random sampling assumption ensures that $f_{W}$ is
identified. Therefore is $f_{W,X^{\ast },X,Y,Z}$\ identified. We can
integrate out $Y$ and/or $Z$ in $f_{W,X^{\ast },X,Y,Z}$\ if they are not
contained in $Q$\ to identify $f_{X^{\ast },Q}$.$\blacksquare $

\bigskip{}

We now consider two parametric models that are most often used in practice
and show how to identify the parameters of interest.

\subsubsection{Linear projection model}

\label{sLPM}

Here $X^{\ast }$ is treated as a cardinal variable. Suppose $X^{\ast }$\ is
observed. Let $\widetilde{Q}=\left( 1,Q^{\top }\right) ^{\top }$.\ We are
interested in $\beta _{C}$, which comes from the following linear projection
model: 
\begin{equation}
X^{\ast }=\widetilde{Q}^{\top }\beta _{C}+\varepsilon ,\text{ where }E\left[ 
\widetilde{Q}\varepsilon \right] =0\text{.}  \label{LP}
\end{equation}%
Then we can identify $\beta _{C}$\ as a least squares solution under
familiar conditions. We state this as a proposition without proof.

\bigskip

\textbf{Proposition 1. }\textit{Suppose Assumption A holds and }$\left(
X^{\ast },\widetilde{Q}\right) $ \textit{satisfies (\ref{LP}). If }$E\left[ 
\widetilde{Q}\widetilde{Q}^{\top }\right] $\textit{\ has full rank, then }%
\begin{equation}
\beta _{C}=\left( E\left[ \widetilde{Q}\widetilde{Q}^{\top }\right] \right)
^{-1}E\left[ \widetilde{Q}X^{\ast }\right] .  \label{bC}
\end{equation}%
\textit{\ }

Note that the linear probability model does not assume a priori that $%
\varepsilon $ is homoskedastic conditional on $\widetilde{Q}$. Here $E\left[ 
\widetilde{Q}\widetilde{Q}^{\top }\right] $ can be identified from the data.

\subsubsection{Ordered probit model}

\label{sOPM}

Now let $X^{\ast }$ be an ordinal variable generated from an ordered
response model. Suppose $X^{\ast }$\ is observed. We are interested in $%
\beta _{O}$, which comes from the following ordered probit model:%
\begin{equation}
X^{\ast }=i\times \mathbf{1}\left[ \mu _{i-1}<\widetilde{Q}^{\top }\beta
_{O}+\sigma \left( Q\right) \varepsilon \leq \mu _{i}\right] \text{ \ for \ }%
i=1,\ldots ,I,  \label{OP}
\end{equation}%
where $\left( \mu _{i}\right) _{i=1}^{I-1}$ is an increasing sequence of
reals with $\mu _{0}=-\infty $ and $\mu _{I}=+\infty $, $\sigma \left(
Q\right) $ denotes a skedastic function that is positive almost surely, and $%
\varepsilon $ has a standard normal distribution.

We can interpret $\widetilde{Q}^{\top }\beta _{O}+\sigma \left( Q\right)
\varepsilon :=U^{\ast }$\ in the traditional way. I.e $U^{\ast }$\ is an
underlying continuous happiness variable that gets transformed into discrete
level of LS. By symmetry of the normal distribution $\widetilde{Q}^{\top
}\beta _{O}$ is the (conditional) median, as well as the mean, of $U^{\ast }$%
. But, unless $\sigma \left( Q\right) =1$ almost surely, the sign of a mean
partial effect of $U^{\ast }$ is generally not identified while the sign of
a median effect is identified. See Section 3.1 of \cite{chen_have_2019} for
a more detailed discussion.

In what follows we denote the CDF of $\varepsilon \ $by $\Phi $. It is
well-known that an ordered probit is not identified and some normalizations
have to be made. In this paper we set $\left( \mu _{1},\mu _{2}\right)
=\left( 0,1\right) $.\footnote{%
Alternatively normalizations can be made on $\beta _{O}$. For example, the
intercept can be set to $0$ and one of the slope parameters can be set to $1$%
.} Next, we show in Lemma 2 that $\sigma $\ is identified without further
assumptions.\ The proof of this result uses the identification strategy from 
\cite{chen_rates_2003}.

\bigskip

\textbf{Lemma 2. }\textit{Suppose Assumption A holds. Then }$\sigma $\textit{%
\ is identified and}%
\begin{equation}
\sigma \left( Q\right) =\frac{1}{\Phi ^{-1}\left( \Pr \left[ X^{\ast }\leq
2|Q\right] \right) -\Phi ^{-1}\left( \Pr \left[ X^{\ast }\leq 1|Q\right]
\right) }.  \label{sigma}
\end{equation}

\textbf{Proof. }From (\ref{OP}), we have:%
\begin{equation}
\Pr \left[ X^{\ast }=i|Q\right] =\Phi \left( \frac{\mu _{i}-\widetilde{Q}%
^{\top }\beta _{O}}{\sigma \left( Q\right) }\right) -\Phi \left( \frac{\mu
_{i-1}-\widetilde{Q}^{\top }\beta _{O}}{\sigma \left( Q\right) }\right)
,\quad i=1,...,I.  \label{Lemma 2o}
\end{equation}%
It then follows that%
\begin{eqnarray}
\Pr \left[ X^{\ast }\leq 1|Q\right] &=&\Phi \left( \frac{-\widetilde{Q}%
^{\top }\beta _{O}}{\sigma \left( Q\right) }\right) ,  \label{Lemma 2i} \\
\Pr \left[ X^{\ast }\leq 2|Q\right] &=&\Phi \left( \frac{1-\widetilde{Q}%
^{\top }\beta _{O}}{\sigma \left( Q\right) }\right) .  \label{Lemma 2ii}
\end{eqnarray}%
So that $\frac{1}{\sigma \left( Q\right) }=\Phi ^{-1}\left( \Pr \left[
X^{\ast }\leq 2|Q\right] \right) -\Phi ^{-1}\left( \Pr \left[ X^{\ast }\leq
1|Q\right] \right) $. By Lemma 1 $f_{X^{\ast }|Q}$\ is identified. Therefore 
$\sigma $\ is identified.$\blacksquare $

\bigskip

An interesting feature of the heteroskedastic ordered response model above
is that we only need information on $\Pr \left[ X^{\ast }=i|Q\right] $ for $%
i=1,2$ to identify $\sigma $\ even if $I$ is larger than $3$. In fact, the
same can be said for the identification of $\beta _{O}$. Suppose that $I\geq
3$, then the additional information from $\Pr \left[ X^{\ast }=i|Q\right] $\
for $i\geq 3$ can be used for identifying $\mu _{O}:=\left( \mu _{3},\ldots
,\mu _{I-1}\right) $.

\bigskip

\textbf{Proposition 2. }\textit{Suppose Assumption A holds and }$\left(
X^{\ast },Q\right) $ \textit{satisfies (\ref{OP}). If }$E\left[ \widetilde{Q}%
\widetilde{Q}^{\top }\right] $\textit{\ has full rank, then}%
\begin{equation}
\beta _{O}=\left( E\left[ \widetilde{Q}\widetilde{Q}^{\top }\right] \right)
^{-1}E\left[ \widetilde{Q}\widetilde{X}^{\ast }\left( Q\right) \right] ,
\label{bO}
\end{equation}%
where $\widetilde{X}^{\ast }\left( Q\right) :=-\sigma \left( Q\right) \Phi
^{-1}\left( \Pr \left[ X^{\ast }=1|Q\right] \right) $ and 
\begin{equation}
\mu _{i}=\widetilde{Q}^{\top }\beta _{O}+\sigma \left( Q\right) \Phi
^{-1}\left( \Pr \left[ X^{\ast }\leq i|Q\right] \right) \text{ \ for \ }%
i=3,\ldots ,I-1\text{.}  \label{mu}
\end{equation}

\textbf{Proof. }Re-arrange (\ref{Lemma 2i}) to obtain, 
\begin{equation*}
-\sigma \left( Q\right) \Phi ^{-1}\left( \Pr \left[ X^{\ast }=1|Q\right]
\right) =\widetilde{Q}^{\top }\beta _{O}.
\end{equation*}%
Pre-multiply both sides of the display above by $\widetilde{Q}$. Take
expectation and solve it to identify $\beta _{O}$.

We can identify $\mu _{O}$ by solving $\Pr \left[ X^{\ast }\leq i|Q\right]
=\Phi \left( \frac{\mu _{i}-\widetilde{Q}^{\top }\beta _{O}}{\sigma \left(
Q\right) }\right) $ for all $i\geq 3$, where the latter expression is
implied by (\ref{Lemma 2o}).$\blacksquare $

\bigskip

By inspecting the proof of Proposition 2, note that we can equivalently use (%
\ref{Lemma 2ii}) to identify $\beta _{O}$. In particular, the normalization
restrictions impose the condition that $\sigma \left( Q\right) \Phi
^{-1}\left( \Pr \left[ X^{\ast }\leq 1|Q\right] \right) =\sigma \left(
Q\right) \Phi ^{-1}\left( \Pr \left[ X^{\ast }\leq 2|Q\right] \right) -1$.

Our discussion above assumes normality of $\varepsilon $\ in (\ref{OP}) for
concreteness. Other parametric models, such as the logit, can be identified
analogously by replacing $\Phi $ with another CDF of a continuous variable
that has full support on $\mathbb{R}$.

\subsection{Practical estimation and inference}

\label{Estimation}

The nonparametric and parametric identification strategies in Section 3.1
and Section 3.2 respectively are constructive. They suggest we can construct
consistent estimators by simply replacing unknown population quantities by
the sample counterparts. But it may not always be ideal to take that
approach in practice. We next provide some alternative estimation methods
for the parameters of interest.

\subsubsection*{Nonparametric estimation}

We can follow the identification steps in Section 3.1\ closely by first
performing an eigen-decomposition using matrices of sample probabilities
instead on the left hand side of equation (\ref{A3}). However, an
eigen-decomposition in finite sample can produce estimates that do not
respect a priori assumed theoretical assumptions. Applications using related
identification results above (see examples in \cite{hu_econometrics_2017})
typically employ a constrained maximum likelihood for estimation.

Under Assumption A, we have 
\begin{eqnarray*}
f_{W,X,Y,Z}\left( w,x,y,z\right) &=&\sum_{x^{\ast }\in \pazocal{X}^{\ast
}}f_{X^{\ast },X,Y,Z|W}\left( x^{\ast },x,y,z|w\right) f_{W}\left( w\right)
\\
&=&\sum_{x^{\ast }\in \pazocal{X}^{\ast }}f_{X|X^{\ast },W}\left( x|x^{\ast
},w\right) f_{Y|X^{\ast },W}\left( y|x^{\ast },w\right) f_{Z|X^{\ast
},W}\left( z|x^{\ast },w\right) f_{X^{\ast }|W}\left( x^{\ast }|w\right)
f_{W}\left( w\right) .
\end{eqnarray*}%
We can therefore construct a likelihood function based on the joint
probability above where the parameters of interest are $\left( f_{X|X^{\ast
},W},f_{Y|X^{\ast },W},f_{Z|X^{\ast },W},f_{X^{\ast }|W},f_{W}\right) $. The
maximum likelihood estimator of $f_{W}$ corresponds to the empirical
distribution of $\left\{ W_{n}\right\} _{n=1}^{N}$, which can be obtained
independently of the other parameters. Maximum likelihood estimation of the
other parameters can be performed conditionally on $W$.

Let $\mathcal{S}_{W},\mathcal{S}_{X},\mathcal{S}_{Y}$ and $\mathcal{S}_{Z}$
denote the cardinalities of the support of $W,X,Y$ and $Z$. Then for each $w$
in the support of $W$, there are $\mathcal{S}_{XYZ}:=\mathcal{S}_{X}\mathcal{%
S}_{Y}\mathcal{S}_{Z}$ possible realizations of $\left( X,Y,Z\right) $.%
\footnote{%
We assume the joint support of $\left( W,X,Y,Z\right) $ is the same for all
realizations of $W$ for notational simplicity.} We can enumerate these
distinct events by $\left\{ x_{j},y_{j},z_{j}\right\} _{j=1}^{\mathcal{S}%
_{XYZ}}$ coupled with $\left\{ m_{j}\right\} _{j=1}^{\mathcal{S}_{XYZ}}$\
where $m_{j}$\ counts how many times realization $j$ occurs in the
sub-sample when $W_{n}=w$. We then estimate the parameters of interest by
maximizing the following conditional\ log-likelihood function 
\begin{equation}
M_{N}\left( \mathbf{p};w\right) =\sum_{j=1}^{\mathcal{S}_{XYZ}}m_{j}\ln
\sum_{x^{\ast }\in \pazocal{X}^{\ast }}p_{X|X^{\ast },W}\left( x_{j}|x^{\ast
},w\right) p_{Y|X^{\ast },W}\left( y_{j}|x^{\ast },w\right) p_{Z|X^{\ast
},W}\left( z_{j}|x^{\ast },w\right) p_{X^{\ast }|W}\left( x^{\ast }|w\right)
,  \label{MLE}
\end{equation}%
where $\mathbf{p}=\left( p_{X|X^{\ast },W},p_{Y|X^{\ast },W},p_{Z|X^{\ast
},W},p_{X^{\ast }|W}\right) $ lies in the parameter space $\mathcal{P}$ that
satisfies the constraints that components of $\mathbf{p}$\ constitute to
valid probability distributions and the inequality relations in Assumption
4. We do this for all $w$ in the support of $W$. Once the nonparametric
estimators of $\left( f_{X|X^{\ast },W},f_{Y|X^{\ast },W},f_{Z|X^{\ast
},W},f_{X^{\ast }|W},f_{W}\right) $ are available, we can proceed to the
parametric estimation stage.

Constrained maximum likelihood estimation is not a computationally simple
task. There are $\mathcal{S}_{X}\left( \mathcal{S}_{X}+\mathcal{S}_{Y}+%
\mathcal{S}_{Z}-3\right) +\mathcal{S}_{X}-1$ free parameters to optimize
over in (\ref{MLE}) for each possible value that $W$ takes.

sub-sample partitioned according to different values of $W$. I.e. we have to
solve this type of optimization problem $\mathcal{S}_{W}$\ times. The
numerical challenge increases with the support size of the variables in the
model. Furthermore, the objective function is not concave so there can be
many local maxima. In practice, we suggest numerical searches should be
performed at different starting points in order to help locate the global
maximum.

\subsubsection*{Parametric Estimation}

Once an estimator for $f_{X^{\ast }|Q}$ is available, population quantities
involving $X^{\ast }$ such as $E\left[ X^{\ast }|Q\right] $ and $\Pr \left[
X^{\ast }\leq i|Q\right] $ can now be estimated even if we do not observe
latent LS. For the linear probability model, from (\ref{bC}), it can be more
convenient to write $\beta _{C}=\left( E\left[ \widetilde{Q}\widetilde{Q}%
^{\top }\right] \right) ^{-1}E\left[ \widetilde{Q}E\left[ X^{\ast }|%
\widetilde{Q}\right] \right] $. We can then estimate $\beta _{C}$\ by
replacing the (unconditional)\ expectation by the sample counterparts.

For the ordered probit model, we can estimate $\sigma $ by replacing $\Pr %
\left[ X^{\ast }\leq i|Q\right] $\ in (\ref{sigma}) by its estimator. Then
we can construct estimators for $\beta _{O}$ and $\mu _{O}$ by replacing the
population moments in (\ref{bO}) and (\ref{mu}) respectively\ by their
sample counterparts. Alternatively, a perhaps more convenient numerical
approach is to estimate the parameters of interest with the build-in
functions of statistical software providing it with the skedastic function
based on (\ref{sigma}). 
\bigskip

\subsubsection*{Inference}

We propose to perform inference by bootstrapping. A bootstrap sample can be
generated by random resampling from the observed data with replacement. The
estimators and tests of nonparametric probabilities and parameters in
Propositions 1 and 2 have regular asymptotic properties that can be
bootstrap as long as the true parameters lie in the interior of the
parameter space (\cite{andrews_estimation_1999, andrews_inconsistency_2000}%
). In practice, estimates of probabilities being close to $0$ or $1$, or any
other a priori (if used) constraints (Assumptions 3 and 4) that appear to be
numerically binding should raise concerns that the assumption of an interior
solution is not being satisfied.

\section{Test for presence of measurement error}

We want to test the hypothesis of no measurement error in LS: 
\begin{equation}
H_{0}^{A}:\Pr \left[ X=X^{\ast }\right] =1.  \label{ME1}
\end{equation}%
Suppose we have $\left( X,Y,Z\right) $\ that satisfies Assumptions 1 - 4.
Then we can identify $f_{X^{\ast },X}$ from $f_{X^{\ast },X,Y,Z}$. One way
to test (\ref{ME1}) directly is to look for evidence that $f_{X^{\ast
},X}\left( x^{\ast },x\right) >0$ for some $x^{\ast }\neq x$. But performing
such test is difficult because the null would imply that $f_{X^{\ast
},X}\left( x^{\ast },x\right) =0$ for all $x^{\ast }\neq x$; parameters at
the boundary will require a non-standard testing procedure. For example, see 
\cite{andrews_testing_2001}. We instead follow the approach of \cite%
{wilhelm_testing_2018}, who shows it is possible to construct a simple test
for the presence of measurement errors under much weaker conditions and
without the need to first identify the entire model.

Theorem 1 in \cite{wilhelm_testing_2018} states that: if $Y\enskip\bot %
\enskip Z\enskip|\enskip X^{\ast }$, then (\ref{ME1}) implies $Y\enskip\bot %
\enskip Z\enskip|\enskip X$. We can then construct a test to detect
potential measurement errors based on a conditional independence hypothesis:%
\begin{equation}
H_{0}^{B}:Y\enskip\bot \enskip Z\enskip|\enskip X.  \label{ME2}
\end{equation}%
We state this as a proposition.

\bigskip

\textbf{Proposition 3. }\textit{Suppose }$Y\enskip\bot \enskip Z\enskip|%
\enskip X^{\ast }$\textit{. Then violation of }$H_{0}^{B}$\textit{\ implies
violation of }$H_{0}^{A}$\textit{.}

\bigskip

The conditional independence assumption in Proposition 3 is already implied
by our Assumption 1. Testing $H_{0}^{B}$\ is just a test of conditional
independence on observed variables. There are many options available for
consistent tests that are easy to construct. In this paper we use a
Kolmogorov-Smirnov type statistic that is based on the sample counterpart of
the following, equivalent, way to write (\ref{ME2}): 
\begin{equation*}
H_{0}^{B}:\max_{\left( x,y,z\right) \in \mathcal{S}_{XYZ}}\left\vert
f_{Y,Z|X}\left( y,z|x\right) -f_{Y|X}\left( y|x\right) f_{Z|X}\left(
z|x\right) \right\vert =0.
\end{equation*}%
In our application we use the frequency estimator for $\left(
f_{Y,Z|X},f_{Y|X},f_{Z|X}\right) $, which corresponds to the maximum
likelihood estimator since $\left( X,Y,Z\right) $ are discrete. Denoting the
frequency estimator by $\left( \widehat{f}_{Y,Z|X},\widehat{f}_{Y|X},%
\widehat{f}_{Z|X}\right) $\bigskip , we have the following test statistic:%
\begin{equation}
TS=\max_{\left( x,y,z\right) \in \mathcal{S}_{XYZ}}\left\vert \widehat{f}%
_{Y,Z|X}\left( y,z|x\right) -\widehat{f}_{Y|X}\left( y|x\right) \widehat{f}%
_{Z|X}\left( z|x\right) \right\vert .  \label{TS}
\end{equation}%
We perform inference by bootstrapping. We construct bootstrap critical
values for $TS$\ from the percentiles of $\left\{ TS^{b}\right\} _{b=1}^{B}$%
, where%
\begin{equation}
TS^{b}=\max_{\left( x,y,z\right) \in \mathcal{S}_{XYZ}}\left\vert \widehat{f}%
_{Y,Z|X}^{b}\left( y,z|x\right) -\widehat{f}_{Y|X}^{b}\left( y|x\right) 
\widehat{f}_{Z|X}^{b}\left( z|x\right) -\left( \widehat{f}_{Y,Z|X}\left(
y,z|x\right) -\widehat{f}_{Y|X}\left( y|x\right) \widehat{f}_{Z|X}\left(
z|x\right) \right) \right\vert ,  \label{TSB}
\end{equation}%
and $\widehat{f}_{A|B}^{b}$ denotes the frequency estimator of $f_{A|B}$\
based on the bootstrap sample. These bootstrap critical values are
consistent as long as $f_{X,Y,Z}$ takes values in the interior of $\left(
0,1\right) $ as discussed at the end of Section 3.\footnote{%
Let $\mathcal{F}\left( x,y,z\right) :=f_{Y,Z|X}\left( y,z|x\right)
-f_{Y|X}\left( y|x\right) f_{Z|X}\left( z|x\right) $ for $\left(
x,y,z\right) \in \mathcal{S}_{XYZ}$. It is clear that $\mathcal{F}\left(
x,y,z\right) $\ is a continuous function of $f_{X,Y,Z}$. Under random
sampling, the asymptotic distribution of $\sqrt{N}\left( \widehat{f}%
_{X,Y,Z}-f_{X,Y,Z}\right) $ can be consistently estimated by $\sqrt{N}\left( 
\widehat{f}_{X,Y,Z}^{b}-\widehat{f}_{X,Y,Z}\right) $ since empirical
measures can be bootstrapped (e.g. see \cite{gin`e_bootstrapping_1990}). The
asymptotic percentiles of $TS$\ can then be consistently estimated using $%
\left\{ TS^{b}\right\} _{b=1}^{B}$ by an application of the Continuous
Mapping Theorem.}

It is worth emphasizing that Proposition 3 only provides a sufficient
condition to detect measurement errors. On the other hand, $H_{0}^{B}$\
generally does not imply $H_{0}^{A}$\ unless additional conditions hold on
the joint distribution of $f_{X^{\ast },X,Y,Z}$. We refer the reader to \cite%
{wilhelm_testing_2018} for further details as to when the two hypotheses are
equivalent.

\section{Application}

We begin this section by describing our dataset and explaining how it is
used in our applications. We report the results of the test for the presence
of measurement error in Section 5.2. We study the effect measurement error
has on a general model of LS in Section 5.3.

\subsection{Data}

\label{datat} We use Wave 3 of the representative household longitudinal
data from UK Understanding Society. The survey covers members of over 35,000
households in the United Kingdom. These data were collected between January
2011 and June 2013. We choose Wave 3 because, unlike the other waves, it
includes questions on personality traits, which is important for us as we
use neuroticism as one of the auxiliary variables for identification.
Further details on the survey questions can be found in Appendix A.

Understanding Society measures LS on a scale from 1 -- 
\enquote{completely
dissatisfied} to 7 -- \enquote{completely satisfied}. For our application,
we aggregate responses to the LS question into 3 larger groups, where 1st
group is those dissatisfied with life overall (%
\enquote{completely
dissatisfied} and \enquote{mostly dissatisfied}), 3rd group is those
satisfied (\enquote{mostly satisfied} and \enquote{completely satisfied})
and the 2nd group is those in between (\enquote{somewhat dissatisfied}, %
\enquote{neither satisfied or dissatisfied} and \enquote{somewhat satisfied}%
). For the GHQ-12 measure, which runs from 0 - \enquote{the least distressed}
to 36 -- \enquote{the most distressed}, we construct $Z$ to share the same
cardinality as $X$ by aggregating all responses below 33rd percentile in
group 1, those between 33rd and 66th percentile in group 2, all the rest in
group 3. The neuroticism score is originally calculated as an average of 3
questions on a scale from 1 to 7. Indicator $Y$ takes the value of 1 if the
level of neuroticism of the individual is above the sample median and zero
otherwise.

We aggregate the data on LS to ensure stable solutions for our constrained
maximum likelihood with both the observed and bootstrap samples. This
reduces the number of parameters to be estimated from 97 to 17 for each
possible realization of the conditioning variables. In particular, we
maximize each of our likelihood function 10 times using a different starting
point to avoid local maxima. Almost all of our estimates converge to the
same solution. On the other hand, if we use a 7-point scale for life
satisfaction we often find numerical optimization starting at a different
point leads to distinct local maxima; in this case we do not have the
confidence that global solutions can be reached in feasible time.

We only use data for the respondents who reported satisfaction with life
overall. This gives us 40,359\ observations from the 49,739 available in the
survey\ (over 81\%). Of those, 56\% are women, 44\% are men. All the
participants are of age 16 or above. 34\% of the respondents have a
long-standing illness or disability, 51.8\% are married, 23.1\% have
obtained a university degree, 5.3\% are unemployed.

Our $W$ consists of: university degree (\textit{degree}), gender (\textit{fem%
}), long-standing illness or disability (\textit{illness}), income above the
sample median (\textit{inc}) and marital status (\textit{married}). Each of
the covariates is a binary variable. That gives $\mathcal{S}_{W}=2^{5}=32$.
We are unable to condition on additional variables because some
socioeconomic groups would have too few observations for nonparametric
estimation. We compute our estimators as described in Section 3.3; in
particular the skedastic function is nonparametric (see (\ref{sigma})).

\subsection{Measurement error in reported LS}

We test for the presence of measurement error in reported LS unconditionally
and conditionally on the covariates. The unconditional test assumes $%
H_{0}^{B}$\ under the null and uses (\ref{TS}) as the test statistic and (%
\ref{TSB}) to construct the critical values. Table \ref{table:521} compares
the value of the test statistic against the bootstrap critical values at
different significant levels. We see very strong evidence against the no
measurement error hypothesis as $H_{0}^{B}$\ is rejected at $1\%$
significance level.

\begin{table}[tbp]
\center
\begin{tabular}{cccc}
\hline\hline
\multirow{2}{*}{TS} & \multicolumn{3}{c}{Critical values} \\ 
\cmidrule(lr){2-4} & \quad 90\% \quad & \quad 95\% \quad & \quad 99\% \quad
\\ \hline
0.106*** & 0.007 & 0.008 & 0.011 \\ \hline\hline
\multicolumn{4}{l}{{\footnotesize {{*} p$<$0.10, {*}{*} p$<$0.05, {*}{*}{*} p%
$<$0.01}}}%
\end{tabular}%
\caption{Unconditional test for the presence of measurement error.}
\label{table:521}
\end{table}

We next look for the presence of measurement error in various socioeconomic
groups. The conditional test partitions the data into $\mathcal{S}_{W}$\
subgroups. In this case, for each $w\in \mathcal{S}_{W}$\ we consider the
following hypothesis:%
\begin{equation*}
H_{0}^{C}\left( w\right) :\max_{\left( x,y,z\right) \in \mathcal{S}%
_{XYZ}}\left\vert f_{Y,Z|X,W}\left( y,z|x,w\right) -f_{Y|X,W}\left(
y|x,w\right) f_{Z|X,W}\left( z|x,w\right) \right\vert =0.
\end{equation*}%
We alter (\ref{TS}) and (\ref{TSB}) to accommodate the conditioning on $W$\
accordingly with the frequency estimator. They are then used respectively to
construct test statistics and bootstrap critical values. Table \ref%
{table:523} gives the test results. The description of different
socioeconomic groups (first column of \ref{table:523}) can be found in
Appendix \ref{aB}.

\begin{table}[tbp]
\center
\begin{tabular}{llcccc}
\hline\hline
\multirow{2}{*}{Group} & \quad \multirow{2}{*}{TS} \quad & 
\multicolumn{3}{c}{Critical values} & \quad \multirow{2}{*}{N} \quad \quad
\\ 
\cmidrule(lr){3-5} &  & \quad 90\% \quad & \quad 95\% \quad & \quad 99\%
\quad &  \\ \hline
0 & 0.114*** & 0.030 & 0.034 & 0.040 & 2,615 \\ 
M & 0.091*** & 0.040 & 0.045 & 0.055 & 1,249 \\ 
I & 0.099*** & 0.037 & 0.042 & 0.052 & 1,832 \\ 
IM & 0.085*** & 0.029 & 0.033 & 0.039 & 3,042 \\ 
H & 0.072*** & 0.031 & 0.035 & 0.042 & 1,153 \\ 
HM & 0.144*** & 0.035 & 0.040 & 0.053 & 1,212 \\ 
HI & 0.087*** & 0.043 & 0.048 & 0.061 & 824 \\ 
HIM & 0.065*** & 0.041 & 0.046 & 0.054 & 1,501 \\ 
F & 0.089*** & 0.025 & 0.028 & 0.034 & 3,627 \\ 
FM & 0.082*** & 0.025 & 0.028 & 0.036 & 3,448 \\ 
FI & 0.082*** & 0.034 & 0.039 & 0.050 & 2,131 \\ 
FIM & 0.079*** & 0.036 & 0.041 & 0.053 & 1,826 \\ 
FH & 0.087*** & 0.023 & 0.025 & 0.031 & 2,092 \\ 
FHM & 0.095*** & 0.028 & 0.032 & 0.041 & 2,278 \\ 
FHI & 0.082*** & 0.030 & 0.034 & 0.041 & 1,373 \\ 
FHIM & 0.083*** & 0.052 & 0.059 & 0.075 & 787 \\ 
D & 0.111** & 0.079 & 0.091 & 0.114 & 327 \\ 
DM & 0.104** & 0.077 & 0.088 & 0.117 & 258 \\ 
DI & 0.102*** & 0.068 & 0.077 & 0.094 & 835 \\ 
DIM & 0.104*** & 0.042 & 0.048 & 0.062 & 1,742 \\ 
DH & 0.064 & 0.093 & 0.102 & 0.127 & 114 \\ 
DHM & 0.045 & 0.064 & 0.072 & 0.088 & 147 \\ 
DHI & 0.129** & 0.088 & 0.098 & 0.129 & 275 \\ 
DHIM & 0.156*** & 0.076 & 0.087 & 0.109 & 632 \\ 
DF & 0.070 & 0.077 & 0.088 & 0.118 & 430 \\ 
DFM & 0.057 & 0.067 & 0.078 & 0.103 & 747 \\ 
DFI & 0.084*** & 0.054 & 0.062 & 0.078 & 1,156 \\ 
DFIM & 0.076*** & 0.052 & 0.059 & 0.073 & 1,360 \\ 
DFH & 0.079** & 0.066 & 0.074 & 0.099 & 196 \\ 
DFHM & 0.058 & 0.110 & 0.129 & 0.164 & 283 \\ 
DFHI & 0.167*** & 0.065 & 0.075 & 0.092 & 424 \\ 
DFHIM & 0.086 & 0.089 & 0.102 & 0.127 & 443 \\ \hline\hline
\multicolumn{6}{l}{{\footnotesize {{*} p$<$0.10, {*}{*} p$<$0.05, {*}{*}{*} p%
$<$0.01}}}%
\end{tabular}%
\caption{Conditional test for the presence of measurement error for
different socioeconomic groups.}
\label{table:523}
\end{table}
We also find very strong evidence that measurement error exists for many
subgroups of the population. In particular, we reject $H_{0}^{C}\left(
w\right) $\ at 1\% significance level for 22 out 32 of socioeconomic groups.
Out of the 10 groups we do not reject the null at 1\%, we reject 4 of them
at 5\%. It is worth noting that the number of observations in these groups
are very small relative to the rest, especially for the groups that we do
not reject the null. The lack of (stronger) evidence to detect measurement
error in some of those groups may be due to small sample size.

\subsection{Estimation results}

Our main results will focus on the distribution of the reported and latent
LS and their ordered probit estimates. In particular, parameter estimates
from probit models are to be interpreted as a component of the conditional
median of the underlying continuous happiness variable. Before we present
them, we consider the effects of reducing the support of reported LS\ from a
7-point scale to a 3-point scale as well as from leaving out some other
covariates. In addition to the variables we have already introduced we will
also use: logarithm of gross personal income (\textit{l\_inc}), unemployment
dummy (\textit{unempl}) and age (\textit{age}) and age squared (\textit{age2}%
).\footnote{%
We need to reduce the support of LS for numerical stability of the maximum
likelihood procedure and limit the number and support of covariates in order
to ensure there is a sufficient number of observations with each
socioeconomic group. For example, from Table \ref{table:523}, we have 10
socioeconomic groups with under 500 observations (with DH the lowest at
117). If we split the sample further with employment status (only 5.3\% are
unemployed) and age bands, there will be groups with too few observations to
estimate 17 parameters.}

Table \ref{table:m1} reports the estimates for the linear projection model
and for the ordered probit model for reported LS with full support (7-point
scale) and reduced support (3-point scale). Here we use personal income
instead of the dummy indicator that a respondent's income is above the
median or not. In this case the heteroskedastic ordered probit is fully
parametric. It is estimated using the oglm STATA command, where the
skedastic function is specified by an exponential function with a linear
index (\cite{williams_fitting_2010}), in order to abstract away from the
need to select tuning parameters from nonparametric estimation (e.g. with
kernel smoothing, see \cite{chen_rates_2003}).

Reducing the support of LS has negligible or no difference in how covariates
affect LS apart from income, where the positive income effect on LS measured
on a 3-point scale is much more pronounced. This pattern holds in all
models. In particular, we note the similarities between the least squares
and the probit estimates are uniform for all covariates (cf. \cite%
{ferrericarbonell_how_2004}) as well as the similarities between results
from homoskedastic and heteroskedastic models (cf. Chen et al. (%
\citeyear{chen_have_2019})\footnote{%
This empirical indifference is in stark contrast to the theoretical
implication illustrated in \citet{bond_sad_nodate}.}). These results are
largely consistent with the literature. Married people are more satisfied
with their lives than their non-married counterparts. Long-standing
illnesses or disability and unemployment significantly reduce LS. Women
report to be more satisfied than men. The effect from age supports the
U-shaped pattern based on a quadratic specification. Money does buy some
happiness. Although there are some conflicted findings on the income effect,
the literature in general seems to find support for the general idea that
income influences LS positively with diminishing returns (e.g. see \cite{clark_relative_2008}). Education is known to influence LS indirectly
through the increase in income and health. A positive effect from having
more education is common result for the studies that cannot fully control
for health\footnote{%
The dataset does not allow us to fully control for the state of health and we only account for the presence of long-standing illness or disability.},
including those for the UK (see, e.g., \citealp{dolan_we_2008}).

\begin{table}[tbp]
\center
\resizebox{\textwidth}{!}{\begin{tabular}{lcccccc}
\hline\hline
& \multicolumn{2}{c}{Linear model} & \multicolumn{2}{c}{Homoskedastic
ordered probit} & \multicolumn{2}{c}{Heteroskedastic ordered probit} \\ 
\cmidrule(lr){2-3} \cmidrule(lr){4-5} \cmidrule(lr){6-7} & Full support  & 
Reduced support & Full support  & Reduced support & Full support  & Reduced support \\ \hline
degree    & 0.212*** & 0.104*** & 0.201*** & 0.163*** & 0.175*** & 0.143*** \\
          & (0.0182)    & (0.0080)     & (0.0216)     & (0.0124)     & (0.0212)  & (0.0165)     \\
fem       & 0.0316**  & 0.0167**  & 0.054*** & 0.027*** & 0.068*** & 0.049*** \\
          & (0.0154)    & (0.0067)     & (0.0180)     & (0.0103)     & (0.0181)     & (0.0120)     \\
health    & -0.488*** & -0.195*** & -0.588*** & -0.300*** & -0.585*** & -0.360*** \\
          & (0.0166)    & (0.0073)     & (0.0196)     & (0.0110)     & (0.0291)     & (0.0256)     \\
mrd       & 0.290*** & 0.124*** & 0.355*** & 0.191*** & 0.375*** & 0.248*** \\
          & (0.0166)     & (0.0073)     & (0.0194)     & (0.0110)     & (0.0233)     & (0.0195)     \\
l\_inc    & 0.0132*   & 0.00899*** & 0.00126      & 0.0144*** & 0.0150*   & 0.0203*** \\
          & (0.00683)   & (0.00299)     & (0.0081)    & (0.0046)     & (0.0089)     & (0.0049)     \\
unempl    & -0.550*** & -0.216*** & -0.546*** & -0.290*** & -0.516*** & -0.267*** \\
          & (0.0373)     & (0.0163)     & (0.0431)     & (0.0237)     & (0.0496)     & (0.0273)     \\
age  & -0.0518*** & -0.0194*** & -0.0662*** & -0.0300*** & -0.0762*** & -0.0435*** \\
          & (0.00251)    & (0.00110)     & (0.00297)     & (0.00170)     & (0.00403)     & (0.00282)     \\
age2 & 0.000599*** & 0.000227*** & 0.000770*** & 0.000356*** & 0.000878*** & 0.000520*** \\
          & (0.0000247)  & (0.0000108)   & (0.0000292) & (0.0000168)   & (0.0000430)     & (0.0000328)     \\
 
\hline\hline
\multicolumn{7}{l}{* p$<$0.1, {**} p$<$0.05, {**}{*}
p$<$0.01}
\end{tabular}}
\caption{Linear projection and ordered probit models estimates for the model
with reported LS (full and reduced support) and full list of covariates.}
\label{table:m1}
\end{table}

\begin{table}[H]
\center
\resizebox{\textwidth}{!}{\begin{tabular}{lcccccc}
\hline\hline
& \multicolumn{2}{c}{Linear model} & \multicolumn{2}{c}{Homoskedastic
ordered probit} & \multicolumn{2}{c}{Heteroskedastic ordered probit} \\ 
\cmidrule(lr){2-3} \cmidrule(lr){4-5} \cmidrule(lr){6-7} & Full support  & 
Reduced support & Full support  & Reduced support & Full support  & Reduced support \\ \hline
degree  & 0.132***  & 0.0710*** & 0.103*** & 0.111*** & 0.0847*** & 0.0803*** \\
        & (0.0183)  & (0.00798) & (0.0213)      & (0.0124)      & (0.0198)      & (0.0146)      \\
fem     & 0.0286*   & 0.0170**   & 0.0393**  & 0.0284*** & 0.0360**  & 0.0495*** \\
        & (0.0153)  & (0.00664) & (0.0177)      & (0.0101)      & (0.0175)      & (0.0118)     \\
illness & -0.409*** & -0.161*** & -0.483*** & -0.245*** & -0.466*** & -0.301*** \\
        & (0.0158)  & (0.0069)  & (0.0183)      & (0.0104)      & (0.0185)      & (0.0114)      \\
income     & 0.0367**   & 0.0293*** & -0.0172  & 0.0435*** & -0.0328*   & 0.0379*** \\
        & (0.0157)  & (0.00683) & (0.0182)      & (0.0104)      & (0.0182)     & (0.0121)      \\
mrd     & 0.244***  & 0.108***  & 0.289*** & 0.170*** & 0.291*** & 0.232*** \\
        & (0.0150)  & (0.00653) & (0.0175)  & (0.0100)      & (0.0175)    & (0.0128)      \\
\hline\hline
\multicolumn{7}{l}{* p$<$0.1, {**} p$<$0.05, {**}{*}
p$<$0.01}
\end{tabular}}
\caption{Linear projection and ordered probit models estimates for the model
with reported LS (full and reduced support) and reduced list of covariates.}
\label{table:m2}
\end{table}
Table \ref{table:m2} contains analogous statistics to Table \ref{table:m1}\
but are based on the reduced set of covariates that we later use to estimate
latent LS. Most of the results between the two tables are qualitatively very
similar. One notable difference, again, is on the income effect. Table \ref%
{table:m2} reports that the income effect remains positive in all cases when
the reduced support LS, but the full support LS yields negative estimates
with the probit model. The negative income effect is, however, weak as it is
insignificant and significant at 10\% in the homoskedastic and
heteroskedastic cases respectively. Our discussion on the income effect from
the previous paragraph applies. Importantly, Table \ref{table:m1}\ and \ref%
{table:m2} suggest that using a median income dummy and omitting
unemployment and age have little impact, as well as reducing the support of LS.

We now provide the estimates from reported and latent LS. Table \ref%
{table:m3} reports the estimates from the linear projection model and the
ordered probit models for reported and latent LS. Here the skedastic
function for the heteroskedastic ordered probit model is estimated
nonparametrically and the normalization is as discussed in Section \ref{sOPM}%
.

\begin{table}[tbp]
\center
\begin{tabular}{lcccccc}
\hline\hline
& \multicolumn{2}{c}{Linear model} & \multicolumn{2}{c}{Homoskedastic
ordered probit} & \multicolumn{2}{c}{Heteroskedastic ordered probit} \\ 
\cmidrule(lr){2-3} \cmidrule(lr){4-5} \cmidrule(lr){6-7} & Reported LS & 
Latent LS & Reported LS & Latent LS & Reported LS & Latent LS \\ \hline
degree & 0.0710*** & -0.0503 & 0.112*** & -0.0754 & 0.0811*** & -0.126 \\ 
& (0.00813) & (0.0491) & (0.0130) & (0.0750) & (0.0148) & (0.0988) \\ 
fem & 0.0170*** & -0.135*** & 0.0284*** & -0.198*** & 0.0505*** & -0.146* \\ 
& (0.00636) & (0.0480) & (0.0096) & (0.0732) & (0.0110) & (0.0826) \\ 
illness & -0.161*** & -0.279*** & -0.245*** & -0.408*** & -0.305*** & 
-0.353*** \\ 
& (0.00703) & (0.0423) & (0.0103) & (0.0645) & (0.0122) & (0.0698) \\ 
income & 0.0293*** & -0.00297 & 0.0436*** & 0.00171 & 0.0368*** & 0.0530 \\ 
& (0.00661) & (0.0423) & (0.0100) & (0.0638) & (0.0119) & (0.0662) \\ 
mrd & 0.108*** & 0.116*** & 0.170*** & 0.169*** & 0.240*** & 0.131** \\ 
& (0.00563) & (0.0386) & (0.00861) & (0.0585) & (0.0105) & (0.0638) \\ 
\hline\hline
\multicolumn{7}{l}{* p$<$0.1, {**} p$<$0.05, {**}{*} p$<$0.01}%
\end{tabular}%
\caption{Linear projection and ordered probit models estimates for the
models with reported and latent LS and reduced list of covariates.}
\label{table:m3}
\end{table}

The results show that the latent LS estimates are qualitatively very similar
across all three models. There is a difference in the signs of the estimates
of income but the income effect is weak and insignificant. Comparing the
results from the models with reported and latent LS, we find the two prominent predictors of LS agree on their effects: health and interpersonal
relationships. People who suffer from long-standing illness or disability
are less satisfied with their life, while married people are more satisfied
than their single counterparts. While the effect of health becomes more
pronounces when we control for the presence of measurement error, it appear
to have substituted the effect on education and income, making them
insignificant with very high p-values. The most striking difference we find
is the gender effect: the female dummy has a positive coefficient for the
reported LS, but negative for the latent one. The same results hold when we
use different constructs of GHQ as a robustness check. See Appendix \ref{aR}.

Our results provide a potential explanation of the \textit{gender puzzle}
based on systematic differences in the misreporting behavior between men and
women. While LS has been widely accepted to be correlated with health and
interpersonal relationship in obvious ways (see, e.g. \cite%
{helliwell_world_2012}), the correlation between well-being and gender
observed in practice is less intuitive. Many surveys find that females
report themselves to be more satisfied with their life than men, e.g. see 
\cite{dolan_we_2008}. These findings are in contradiction with being
worse off in many measurable social and economic outcomes, which are known
to be the sources of well-being (pay gap and unemployment gap, to name a
few). More recently, \cite{meisenberg_gender_2015} use a
dataset of 90 countries represented in the World Values Survey to find that
gender equality, gainful employment and prolonged schooling decrease female
well-being; while women are happier in the countries that maintain
traditional gender roles. 

In order to better understand the difference in reporting behavior of men
and women, we consider 4 particular socioeconomic groups of respondents
presented in Table \ref{table:62}. Group O contains single men with income
below the median, with no degree and no long-standing illness or disability.
Group H contains the same respondents who suffer from long standing health
issues. The other two groups are the female respondents with the same
characteristics. Distributions for the other groups are presented in
Appendix \ref{aD}. 
\begin{table}[tbp]
\centering
\begin{tabular}{lcccc}
\hline\hline
& \quad 0 & \quad H & \quad F & \quad FH\quad \\ \hline
fem & \quad 0 & \quad 0 & \quad 1 & \quad 1 \quad \\ 
mrd & \quad 0 & \quad 0 & \quad 0 & \quad 0 \quad \\ 
degree & \quad 0 & \quad 0 & \quad 0 & \quad 0 \quad \\ 
illness & \quad 0 & \quad 1 & \quad 0 & \quad 1 \quad \\ 
inc & \quad 0 & \quad 0 & \quad 0 & \quad 0 \quad \\ \hline\hline
\end{tabular}%
\caption{Characteristics of the groups.}
\label{table:62}
\end{table}

\begin{table}[]
\centering
\begin{tabular}{cc}
\quad\quad\quad\quad 0 & \quad\quad\quad\quad H \\ 
$\quad\mathbf{M}_{X}=%
\begin{bmatrix}
0.0887 & 0.3606 & 0.5507 \\ 
(0.0054) & (0.0083) & (0.0088)%
\end{bmatrix}%
$ & $\quad\mathbf{M}_{X}=%
\begin{bmatrix}
0.1657 & 0.4709 & 0.3634 \\ 
(0.0100) & (0.0151) & (0.0142)%
\end{bmatrix}%
$ \\ 
&  \\ 
$\mathbf{M}_{X|X^{*}}=%
\begin{bmatrix}
0.2395 & 0.0617 & 0.0720 \\ 
(0.0518) & (0.0109) & (0.0082) \\ 
0.6915 & 0.4945 & 0.1460 \\ 
(0.0379) & (0.0382) & (0.0216) \\ 
0.0691 & 0.4437 & 0.7819 \\ 
(0.0614) & (0.0366) & (0.0223)%
\end{bmatrix}%
$ & $\mathbf{M}_{X|X^{*}}=%
\begin{bmatrix}
0.4562 & 0.0554 & 0.0562 \\ 
(0.0656) & (0.0325) & (0.0192) \\ 
0.5438 & 0.5899 & 0.2541 \\ 
(0.0641) & (0.0609) & (0.0435) \\ 
0.0001 & 0.3547 & 0.6898 \\ 
(0.0177) & (0.0712) & (0.0454)%
\end{bmatrix}%
$ \\ 
&  \\ 
$\quad\mathbf{M}_{X^{*}}=%
\begin{bmatrix}
0.1254 & 0.4195 & 0.4551 \\ 
(0.0347) & (0.0516) & (0.0459)%
\end{bmatrix}%
$ & $\quad\mathbf{M}_{X^{*}}=%
\begin{bmatrix}
0.2745 & 0.4090 & 0.3165 \\ 
(0.0493) & (0.0530) & (0.0519)%
\end{bmatrix}%
$ \\ 
&  \\ 
\quad\quad\quad\quad F & \quad\quad\quad\quad FH \\ 
$\quad\mathbf{M}_{X}=%
\begin{bmatrix}
0.0940 & 0.3226 & 0.5834 \\ 
(0.0050) & (0.0076) & (0.0076)%
\end{bmatrix}%
$ & $\quad\mathbf{M}_{X}=%
\begin{bmatrix}
0.1472 & 0.4402 & 0.4125 \\ 
(0.0078) & (0.0109) & (0.0105)%
\end{bmatrix}%
$ \\ 
&  \\ 
$\mathbf{M}_{X|X^{*}}=%
\begin{bmatrix}
0.1525 & 0.0649 & 0.0751 \\ 
(0.0270) & (0.0092) & (0.0096) \\ 
0.5947 & 0.3794 & 0.0902 \\ 
(0.0402) & (0.0350) & (0.0179) \\ 
0.2528 & 0.5557 & 0.8346 \\ 
(0.0596) & (0.0360) & (0.0179)%
\end{bmatrix}%
$ & $\mathbf{M}_{X|X^{*}}=%
\begin{bmatrix}
0.2662 & 0.0882 & 0.0448 \\ 
(0.0284) & (0.0187) & (0.0163) \\ 
0.6216 & 0.4497 & 0.0924 \\ 
(0.0258) & (0.0434) & (0.0342) \\ 
0.1121 & 0.4621 & 0.8628 \\ 
(0.0382) & (0.0523) & (0.0384)%
\end{bmatrix}%
$ \\ 
&  \\ 
$\quad\mathbf{M}_{X^{*}}=%
\begin{bmatrix}
0.2847 & 0.3068 & 0.4085 \\ 
(0.0460) & (0.0588) & (0.0403)%
\end{bmatrix}%
$ & $\quad\mathbf{M}_{X^{*}}=%
\begin{bmatrix}
0.3831 & 0.4061 & 0.2108 \\ 
(0.0562) & (0.0440) & (0.0421)%
\end{bmatrix}%
$ \\ 
& 
\end{tabular}%
\caption{Distribution of reported and latent LS.}
\label{table:63}
\end{table}


Comparing the upper (0 and H) and the lower (F and FH) blocks of Table \ref%
{table:63} explains the different signs of the gender dummy coefficients in
two models. The distribution of reported LS, $\mathbf{M}_{X}$, is similar
for men (upper block) and women (lower block) with women slightly more
likely to report the high state, hence positive coefficient for the gender
dummy. The comparison of latent distributions, $\mathbf{M}_{X^{\ast }}$,
shows the opposite: women are more likely to be in the low state and less
likely to be in the high one. However, we do not observe lower levels of LS
among women in the data, because they misreport in a systematically
different way compared to men. Comparing the matrices of misreporting
probabilities, $\mathbf{M}_{X|X^{\ast }}$, shows that though all the
respondents are prone to report higher states that they latently are, women
do it more emphatically. I.e. women are more likely than men to report the
highest state, regardless of their latent state.

While our econometric analysis cannot provide a behavioral answer to this
finding, it helps identify differences in reporting behavior for further
research to be done to rationalize them. At the moment we can only offer
potential explanations. One particular conjecture is based on the distinct
patterns of {happy image management} induced by the influence of gender
roles and social stereotypes. \cite{kahneman_well-being:_1999} and
references therein suggest that according to traditional gender roles women
are usually seen as more cheerful and enthusiastic. As a result women might
report higher states conforming with the existing norm.

\section{Conclusion}

There is an enormous interest in using subjective well-being data in
economics and related disciplines. Existing research almost always ignores
measurement error despite the fact that the literature acknowledges it is
expected to be present. In particular, the error is non-classical and its
potential effects on subsequent analysis is completely unknown. In this
paper we use novel nonparametric techniques to formally test for the
presence of measurement error and empirically investigate its effects.

Our test is based on the idea proposed in \cite{wilhelm_testing_2018} and we
use a misclassification model of \citet{hu_identification_2008}. The
application of these nonparametric methods in itself is not entirely
trivial. Primarily there is an empirical challenge in finding appropriate
data that satisfies assumptions somewhat analogous to finding a good
instrument. We use Wave 3 of the UK Understanding Society survey because it
is the only wave that contains questions on neuroticism that we believe is
crucial for our empirical study. We also have to convert existing
nonparametric identification results into parametric ones because a fully
nonparametric approach has a limited scope in practice as it does not allow
to include many covariates, not to mention that all the benchmark results in
the literature are parametric. For a parametric estimation, we in particular
advocate the use of a heteroskedastic ordered response model to analyze
wellbeing data that builds on the argument of \cite{chen_have_2019} to avoid
recent critiques highlighted in \citet{bond_sad_nodate}.

We find evidence of measurement error in LS for the whole sample as well as
26 out of 32 socioeconomic subgroups of the data. We use covariates that
define these subgroups to estimate a model of LS. We find the most important
drivers of LS are the same in both models of latent LS and reported LS. The
most notable difference is the gender effect. The happiness literature often
finds women reporting higher levels of well-being, despite being worse of in
measurable objective outcomes, e.g. income, employment, etc. This puzzle can
be rationalized by our model because women are more likely to report
themselves to be happier than they actually are compared to men.

The puzzling relations between female well-being and socioeconomic measures
were also found in the panel data. \cite{stevenson_paradox_2009} show that
reported well-being of women in the United States declined in the last 35
years despite the improvement of women's positions in many objective
outcomes. The authors label this result as \textit{The Paradox of Declining
Female Happiness}. In order to further investigate whether the gender puzzle
can be explained by measurement error, we need to have panel data to extend
our analysis. 

One methodological recommendation of our research is for future surveys to
consider collecting data that increase the scope to apply modern econometric
techniques to solve old problems like measurement error. For example, the UK
Understanding Society data is in fact longitudinal. But the lack of
information on neuroticism from all Waves other than Wave 3 prevents us from
controlling from individual specific effects that would be very helpful in
well-being studies.

\newpage{}

\appendix

\section{Data appendix}

\label{data}

\begin{figure}[H]
\centering \includegraphics[width=0.7\linewidth]{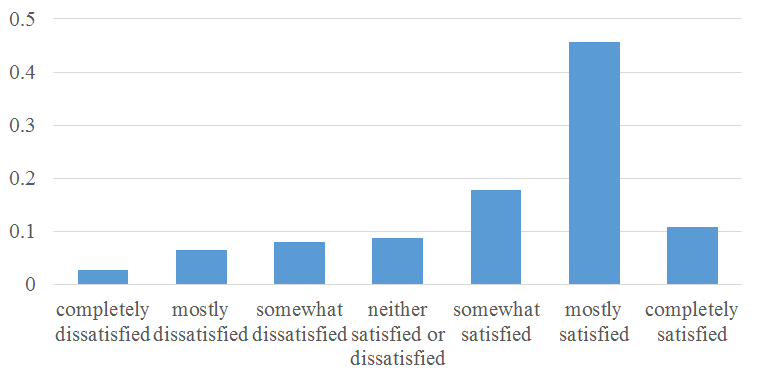}
\caption{Distribution of reported LS.}
\label{fig.s}
\end{figure}

We use reported LS as a possibly mismeasured counterpart of the latent LS,
which is our variable of interest. Respondents report satisfaction with life
overall as a part of a Self-Completion Satisfaction module. They are asked
to choose the number which they feel best describes how dissatisfied or
satisfied they are with their life overall. Scale goes from 1 to 7 and
includes 1 -- {completely dissatisfied}, 2 -- {mostly dissatisfied}, 3 -- {%
somewhat dissatisfied}, 4 -- {neither satisfied or dissatisfied}, 5 -- {%
somewhat satisfied}, 6 -- {mostly satisfied}, and 7 -- {completely satisfied}%
. Out of all the respondents who reported satisfaction with life overall,
45.5\% are {mostly satisfied}, 17.8\% are {somewhat satisfied}, only 9.5\%
of respondents are reportedly {completely dissatisfied} or {mostly
dissatisfied}. The distribution of the self-reported LS is presented in
Figure \ref{fig.s}.

To estimate the parameters of interest for the latent LS instead of the
reported one we need to chose two other related measures or auxiliary
variables. Both variables $Z$ and $Y$ are not required to be measured
without error, however, they should be free of the measurement error we wish
to control for.

The perfect $Z$ and $Y$ would be those strongly correlated with latent life
satisfaction and measured with error that is independent of error occurring
in surveys, e.g. coming from administrative data. However, data like that is
not available as well-being cannot be measured in an administrative way, so
we are constrained to data collected via surveys.

The variables we chose are subjective well-being from the General Health
Questionnaire (GHQ) as a second measure $Z$ and the level of neuroticism for
an indicator $Y$. The measure of well-being is a natural candidate for the
second measure of $X^{*}$, as it portrays a similar concept. Neuroticism is
a strong predictor of well-being as shown by multiple studies. The variables
are correlated with reported satisfaction with life overall (Table \ref%
{table:a21}).

\begin{table}[]
\par
\begin{centering}
\begin{tabular}{lccc}
\hline 
\hline
 & Satisfaction with  & Well-being (GHQ)  & Neuroticism \\
 & life overall  &  & \\
\hline 
Satisfaction with life overall  & 1.0000  &  & \\
Well-being (GHQ)  & -0.4535  & 1.0000  & \\
Neuroticism  & -0.2564  & 0.4837  & 1.0000 \\
\hline \hline
\end{tabular}
\par\end{centering}
\caption{Covariance matrix for reported LS, wellbeing and neuroticism.}
\label{table:a21}
\end{table}

The Understanding Society dataset provides a measure of mental state which
is derived from 12 questions of General Health Questionnaire. Here is the
full list of the GHQ questions:

\begin{itemize}
\item The next questions are about how you have been feeling recently.

\begin{itemize}
\item Have you recently been able to concentrate on whatever you're doing?

\item Have you recently lost much sleep over worry?

\item Have you recently felt that you were playing a useful part in things?

\item Have you recently felt capable of making decisions about things?

\item Have you recently felt constantly under strain?

\item Have you recently felt you couldn't overcome your difficulties?

\item Have you recently been able to enjoy your normal day-to-day activities?

\item Have you recently been able to face up to problems?

\item Have you recently been feeling unhappy or depressed?

\item Have you recently been losing confidence in yourself?

\item Have you recently been thinking of yourself as a worthless person?

\item Have you recently been feeling reasonably happy, all things considered?
\end{itemize}
\end{itemize}

\begin{figure}[H]
\centering \includegraphics[width=0.7\linewidth]{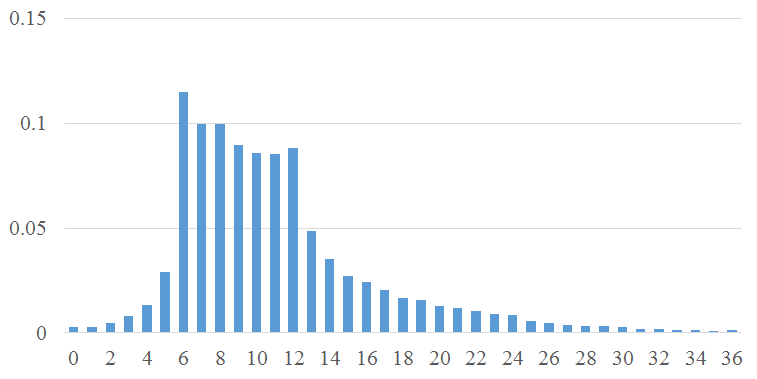}
\caption{Distribution of subjective well-being (GHQ).}
\label{fig.swb}
\end{figure}

The questions are a part of the Self-Completion Module. We now describe how
the measure is constructed. Respondents choose one the following answers: 1
-- {better than usual}, 2 -- {same as usual}, 3 -- {less than usual} and 4
-- {much less than usual}. Then valid answers are converted to a single
scale by re-coding so that the scale for individual variables runs from 0 to
3 instead of 1 to 4, and then summing up. That provides a scale running from
0 (the least distressed) to 36 (the most distressed). The distribution of
the measure is presented in Figure \ref{fig.swb}.

We use a measure of neuroticism for an indicator variable $Y$. Neuroticism
is one of the Big Five personality traits. The other 4 variables are
agreeableness, conscientiousness, extroversion and openness. Wave 3 of
Understanding Society is the only wave that measures the Big Five
personality traits and that is why we use this wave for the analysis. Each
personality trait is derived as an average of the answers to the three
questions. For the measure of neuroticism, respondents are asked to report
on the scale from 1 -- {doesn't apply to me at all} to 7 -- {applies to me
perfectly}, their opinion of the following statements:

\begin{itemize}
\item How much you can relate to the following statements:

\begin{itemize}
\item I see myself as someone who worries a lot;

\item I see myself as someone who is nervous;

\item I see myself as someone who is relaxed;
\end{itemize}
\end{itemize}

\begin{figure}[H]
\centering \includegraphics[width=0.7\linewidth]{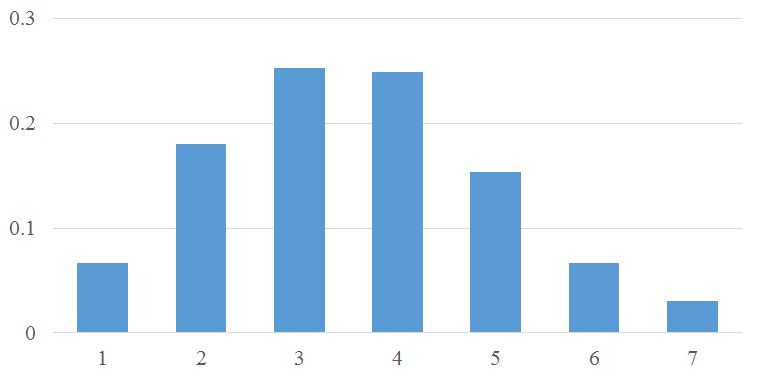}
\caption{Distribution of the level of neuroticism.}
\label{fig.n}
\end{figure}

Quarter of all respondents report an average of 3, less that 10\% report 6
or 7 (Figure \ref{fig.n}).

\newpage

\section{Description of the subgroups for the model of LS}

\label{aB} 
\begin{table}[H]
\centering
\resizebox{0.55\textwidth}{!}{\begin{tabular}{lccccc}
\hline\hline
& degree & female & health & income & married \\ \hline
0 & 0 & 0 & 0 & 0 & 0 \\ 
M & 0 & 0 & 0 & 0 & 1 \\ 
I & 0 & 0 & 0 & 1 & 0 \\ 
IM & 0 & 0 & 0 & 1 & 1 \\ 
H & 0 & 0 & 1 & 0 & 0 \\ 
HM & 0 & 0 & 1 & 0 & 1 \\ 
HI & 0 & 0 & 1 & 1 & 0 \\ 
HIM & 0 & 0 & 1 & 1 & 1 \\ 
F & 0 & 1 & 0 & 0 & 0 \\ 
FM & 0 & 1 & 0 & 0 & 1 \\ 
FI & 0 & 1 & 0 & 1 & 0 \\ 
FIM & 0 & 1 & 0 & 1 & 1 \\ 
FH & 0 & 1 & 1 & 0 & 0 \\ 
FHM & 0 & 1 & 1 & 0 & 1 \\ 
FHI & 0 & 1 & 1 & 1 & 0 \\ 
FHIM & 0 & 1 & 1 & 1 & 1 \\ 
D & 1 & 0 & 0 & 0 & 0 \\ 
DM & 1 & 0 & 0 & 0 & 1 \\ 
DI & 1 & 0 & 0 & 1 & 0 \\ 
DIM & 1 & 0 & 0 & 1 & 1 \\ 
DH & 1 & 0 & 1 & 0 & 0 \\ 
DHM & 1 & 0 & 1 & 0 & 1 \\ 
DHI & 1 & 0 & 1 & 1 & 0 \\ 
DHIM & 1 & 0 & 1 & 1 & 1 \\ 
DF & 1 & 1 & 0 & 0 & 0 \\ 
DFM & 1 & 1 & 0 & 0 & 1 \\ 
DFI & 1 & 1 & 0 & 1 & 0 \\ 
DFIM & 1 & 1 & 0 & 1 & 1 \\ 
DFH & 1 & 1 & 1 & 0 & 0 \\ 
DFHM & 1 & 1 & 1 & 0 & 1 \\ 
DFHI & 1 & 1 & 1 & 1 & 0 \\ 
DFHIM & 1 & 1 & 1 & 1 & 1 \\ \hline\hline
\end{tabular}}
\caption{Description of the socioeconomic subgroups.}
\label{table:A41}
\end{table}

\section{Robustness checks}

\label{aR}

Among all the assumptions required for identification, the assumption that
is the least trivial to satisfy is the Conditional Independence Assumption
discussed in Section \ref{Identification}. Since it is imposed on the
unobserved variable we cannot directly test it from data. The assumption
would be violated if the reported LS, $X$, and one of the auxiliary
variables, $Z$ or $Y$, contain an error from the same source. $Y$ is
constructed from the questions about one's behavior and personal traits,
thus it is unlikely to contain the same error as a LS question. GHQ-12, on
the other hand, is an alternative measure of well-being, $Z$. Since it
includes subjective questions, for example questions about feeling
reasonably happy, it might be the case that the same error would enter both
variables. To ensure that the results are not driven by the presence of
similar error, we estimate the parameters of interest for alternative
candidates for $Z$.

We divide GHQ-12 into two subgroups, one of those includes questions which
are relatively less subjective (GHQ1), another one contains questions which
are relatively more subjective and closer to the LS question (GHQ2). 
Despite some differences in the results, the main findings remain unchanged.
We further describe the variables used.

\begin{enumerate}
\item GHQ questions, which are divided into 2 groups (each question is
measured on the scale from 1 to 4)

\begin{enumerate}
\item GHQ1

\begin{itemize}
\item Have you recently been able to concentrate on whatever you're doing?

\item Have you recently lost much sleep over worry?

\item Have you recently felt capable of making decisions about things?

\item Have you recently felt you couldn't overcome your difficulties?

\item Have you recently been able to face up to problems?

\item Have you recently been losing confidence in yourself?
\end{itemize}

\item GHQ2

\begin{itemize}
\item Have you recently felt that you were playing a useful part in things?

\item Have you recently felt constantly under strain?

\item Have you recently been able to enjoy your normal day-to-day activities?

\item Have you recently been feeling unhappy or depressed?

\item Have you recently been thinking of yourself as a worthless person?

\item Have you recently been feeling reasonably happy, all things considered?
\end{itemize}
\end{enumerate}


\end{enumerate}

Table \ref{table:14} contains the covariance matrix for the different
instruments. Tables \ref{table:A31} and \ref{table:A32} provide the estimates
of the linear projection and probit models.\newline

\begin{table}[H]
\centering
\begin{tabular}{l|cccccc}
\hline\hline
& LS & Neuroticism & GHQ & GHQ1 & GHQ2 &  \\ \hline
LS & 1.0000 &  &  &  &  &  \\ 
Neuroticism & -0.2563 & 1.0000 &  &  &  &  \\ 
GHQ & -0.4533 & 0.4837 & 1.0000 &  &  &  \\ 
GHQ1 & -0.4225 & 0.4796 & 0.9613 & 1.0000 &  &  \\ 
GHQ2 & -0.4504 & 0.4537 & 0.9661 & 0.8577 & 1.0000 &  \\ \hline\hline
\end{tabular}
\caption{Covariance matrix for different choices of instruments.}
\label{table:14}
\end{table}

\begin{table}[H]
\centering
\resizebox{0.6\textwidth}{!}{\begin{tabular}{lcccc}
\hline\hline
& Reported LS & Latent LS & Latent LS & Latent LS \\ 
&  & GHQ & GHQ1 & GHQ2 \\ \hline
degree & 0.071 & -0.050 & -0.068 & -0.089 \\ 
& (0.0081) & (0.0491) & (0.0713) & (0.0569) \\ 
fem & 0.017 & -0.135 & -0.199 & -0.185 \\ 
& (0.0064) & (0.0480) & (0.0587) & (0.0573) \\ 
illness & -0.161 & -0.279 & -0.172 & -0.197 \\ 
& (0.0070) & (0.0423) & (0.0543) & (0.0500) \\ 
income & 0.029 & -0.003 & 0.026 & 0.048 \\ 
& (0.0066) & (0.0423) & (0.0575) & (0.0566) \\ 
mrd & 0.108 & 0.116 & 0.105 & 0.156 \\ 
& (0.0056) & (0.0386) & (0.0621) & (0.0518) \\ \hline\hline
\end{tabular}}%
\caption{Linear projection estimates for LS model.}
\label{table:A31}
\end{table}

\begin{table}[H]
\centering
\resizebox{\textwidth}{!}{\begin{tabular}{lcccccccc}
\hline
\hline
        & \multicolumn{4}{c}{Homoskedastic ordered probit}          & \multicolumn{4}{c}{Heteroskedastic ordered probit}  \\ \cmidrule(lr){2-5} \cmidrule(lr){6-9}
        & Reported LS & Latent LS & Latent LS              & Latent LS & Reported LS & Latent LS & Latent LS                & Latent LS             \\
        &             & GHQ       & GHQ1                   & GHQ2      &             & GHQ       & GHQ1                     & GHQ2                  \\
          \hline
degree  & 0.112       & -0.075    & -0.088                 & -0.124    & 0.081       & -0.126    & -0.109                   & -0.155                \\
        & (0.0130)    & (0.0750)  & (0.0929)               & (0.0790)  & (0.0148)    & (0.0988)  & (0.1067)                 & (0.1037)              \\
fem     & 0.028       & -0.198    & -0.260                 & -0.259    & 0.050       & -0.146    & -0.260                   & -0.253                \\
        & (0.0096)    & (0.0732)  & (0.0761)               & (0.0792)  & (0.0110)    & (0.0826)  & (0.0958)                 & (0.0874)              \\
illness & -0.245      & -0.408    & -0.224                 & -0.269    & -0.305      & -0.353    & -0.194                   & -0.247                \\
        & (0.0103)    & (0.0645)  & (0.0693)               & (0.0673)  & (0.0122)    & (0.0698)  & (0.0859)                 & (0.0767)              \\
income  & 0.044       & 0.002     & 0.035                  & 0.073     & 0.037       & 0.053     & 0.048                    & 0.129                 \\
        & (0.0100)    & (0.0638)  & (0.0737)               & (0.0786)  & (0.0119)    & (0.0662)  & (0.0856)                 & (0.1053)              \\
mrd     & 0.170       & 0.169     & 0.136                  & 0.215     & 0.240       & 0.131     & 0.156                    & 0.174                 \\
        & (0.0086)    & (0.0585)  & (0.0791)               & (0.0710)  & (0.0105)    & (0.0638)  & (0.0899)                 & (0.0831)              \\
        \hline
        \hline
\end{tabular}}
\caption{Ordered probit estimates for LS model.}
\label{table:A32}
\end{table}

\newpage
\section{Distribution of reported and latent LS}

\label{aD}

\begin{table}[H]
\resizebox{\textwidth}{!}{\begin{tabular}{ccc}
\quad\quad\quad\quad 0 & \quad\quad\quad\quad M & \quad\quad\quad\quad I \\ 
﻿$\quad \mathbf{M}_{X}=
\begin{bmatrix}   0.0887& 0.3606& 0.5507\\(0.0054)& (0.0083)& (0.0088) \end{bmatrix}$ &
﻿$\quad \mathbf{M}_{X}=
\begin{bmatrix}   0.0977& 0.2922& 0.6101\\(0.0086)& (0.0121)& (0.0121) \end{bmatrix}$ & 
﻿$\quad \mathbf{M}_{X}=
\begin{bmatrix}   0.0721& 0.3684& 0.5595\\(0.0061)& (0.0108)& (0.0113) \end{bmatrix}$  \\& & \\ 
$\mathbf{M}_{X|X^{*}}=
\begin{bmatrix}   0.2395& 0.0617& 0.0720\\(0.0518)& (0.0109)& (0.0082)\\0.6915& 0.4945& 0.1460\\(0.0379)& (0.0382)& (0.0216)\\0.0691& 0.4437& 0.7819\\(0.0614)& (0.0366)& (0.0223) \end{bmatrix}$ &
$\mathbf{M}_{X|X^{*}}=
\begin{bmatrix}   0.1276& 0.0875& 0.0926\\(0.0495)& (0.0619)& (0.0180)\\0.6737& 0.2748& 0.0429\\(0.0986)& (0.1288)& (0.0308)\\0.1987& 0.6377& 0.8645\\(0.0908)& (0.1155)& (0.0282) \end{bmatrix}$ &
$\mathbf{M}_{X|X^{*}}=
\begin{bmatrix}   0.1256& 0.0510& 0.0679\\(0.0214)& (0.0224)& (0.0158)\\0.6986& 0.4609& 0.1851\\(0.0581)& (0.0811)& (0.0299)\\0.1758& 0.4881& 0.7470\\(0.0626)& (0.0932)& (0.0272) \end{bmatrix}$  \\& & \\ 
$\quad\mathbf{M}_{X^{*}}=
\begin{bmatrix}   0.1254& 0.4195& 0.4551\\(0.0347)& (0.0516)& (0.0459) \end{bmatrix}$ &
$\quad\mathbf{M}_{X^{*}}=
\begin{bmatrix}   0.2160& 0.4875& 0.2965\\(0.0633)& (0.1141)& (0.1019) \end{bmatrix}$ &
$\quad\mathbf{M}_{X^{*}}=
\begin{bmatrix}   0.1725& 0.3435& 0.4840\\(0.0444)& (0.0772)& (0.0847) \end{bmatrix}$  \\& & \\ 

\quad\quad\quad\quad IM & \quad\quad\quad\quad H & \quad\quad\quad\quad HM \\ 
﻿$\quad \mathbf{M}_{X}=
\begin{bmatrix}   0.0832& 0.2847& 0.6321\\(0.0047)& (0.0076)& (0.0078) \end{bmatrix}$ &
﻿$\quad \mathbf{M}_{X}=
\begin{bmatrix}   0.1657& 0.4709& 0.3634\\(0.0100)& (0.0151)& (0.0142) \end{bmatrix}$ & 
﻿$\quad \mathbf{M}_{X}=
\begin{bmatrix}   0.1114& 0.4092& 0.4794\\(0.0088)& (0.0128)& (0.0137) \end{bmatrix}$  \\& & \\ 
$\mathbf{M}_{X|X^{*}}=
\begin{bmatrix}   0.1347& 0.0841& 0.0766\\(0.0471)& (0.0935)& (0.0081)\\0.7392& 0.4908& 0.1251\\(0.1026)& (0.1690)& (0.0253)\\0.1261& 0.4251& 0.7983\\(0.1057)& (0.1463)& (0.0270) \end{bmatrix}$ &
$\mathbf{M}_{X|X^{*}}=
\begin{bmatrix}   0.4562& 0.0554& 0.0562\\(0.0656)& (0.0325)& (0.0192)\\0.5438& 0.5899& 0.2541\\(0.0641)& (0.0609)& (0.0435)\\0.0001& 0.3547& 0.6898\\(0.0177)& (0.0712)& (0.0454) \end{bmatrix}$ &
$\mathbf{M}_{X|X^{*}}=
\begin{bmatrix}   0.3481& 0.0344& 0.0723\\(0.0409)& (0.0137)& (0.0147)\\0.6503& 0.5005& 0.1956\\(0.0400)& (0.0482)& (0.0229)\\0.0017& 0.4651& 0.7321\\(0.0245)& (0.0476)& (0.0237) \end{bmatrix}$  \\& & \\ 
$\quad\mathbf{M}_{X^{*}}=
\begin{bmatrix}   0.0718& 0.3159& 0.6124\\(0.0764)& (0.0933)& (0.0722) \end{bmatrix}$ &
$\quad\mathbf{M}_{X^{*}}=
\begin{bmatrix}   0.2745& 0.4090& 0.3165\\(0.0493)& (0.0530)& (0.0519) \end{bmatrix}$ &
$\quad\mathbf{M}_{X^{*}}=
\begin{bmatrix}   0.1976& 0.4060& 0.3964\\(0.0244)& (0.0443)& (0.0502) \end{bmatrix}$  \\& & \\ 

\quad\quad\quad\quad HI & \quad\quad\quad\quad HIM & \quad\quad\quad\quad F \\ 
﻿$\quad \mathbf{M}_{X}=
\begin{bmatrix}   0.1177& 0.4769& 0.4053\\(0.0108)& (0.0160)& (0.0170) \end{bmatrix}$ &
﻿$\quad \mathbf{M}_{X}=
\begin{bmatrix}   0.0906& 0.3684& 0.5410\\(0.0068)& (0.0120)& (0.0122) \end{bmatrix}$ & 
﻿$\quad \mathbf{M}_{X}=
\begin{bmatrix}   0.0940& 0.3226& 0.5834\\(0.0050)& (0.0076)& (0.0076) \end{bmatrix}$  \\& & \\ 
$\mathbf{M}_{X|X^{*}}=
\begin{bmatrix}   0.2307& 0.0000& 0.0821\\(0.0771)& (0.2378)& (0.0343)\\0.6775& 0.6136& 0.1809\\(0.0678)& (0.2286)& (0.0665)\\0.0918& 0.3864& 0.7370\\(0.0318)& (0.1290)& (0.0475) \end{bmatrix}$ &
$\mathbf{M}_{X|X^{*}}=
\begin{bmatrix}   0.1592& 0.0645& 0.0566\\(0.0376)& (0.1047)& (0.0149)\\0.7010& 0.3726& 0.1307\\(0.0612)& (0.1317)& (0.0330)\\0.1398& 0.5629& 0.8127\\(0.0622)& (0.1329)& (0.0297) \end{bmatrix}$ &
$\mathbf{M}_{X|X^{*}}=
\begin{bmatrix}   0.1525& 0.0649& 0.0751\\(0.0270)& (0.0092)& (0.0096)\\0.5947& 0.3794& 0.0902\\(0.0402)& (0.0350)& (0.0179)\\0.2528& 0.5557& 0.8346\\(0.0596)& (0.0360)& (0.0179) \end{bmatrix}$  \\& & \\ 
$\quad\mathbf{M}_{X^{*}}=
\begin{bmatrix}   0.3779& 0.2505& 0.3716\\(0.0534)& (0.0774)& (0.0618) \end{bmatrix}$ &
$\quad\mathbf{M}_{X^{*}}=
\begin{bmatrix}   0.3126& 0.2457& 0.4417\\(0.0700)& (0.0923)& (0.0930) \end{bmatrix}$ &
$\quad\mathbf{M}_{X^{*}}=
\begin{bmatrix}   0.2847& 0.3068& 0.4085\\(0.0460)& (0.0588)& (0.0403) \end{bmatrix}$  \\& & \\ 
\end{tabular}
}
\caption{Distribution of reported and latent LS for different socioeconomic
groups.}
\label{table:73}
\end{table}

\begin{table}[H]
\resizebox{\textwidth}{!}{\begin{tabular}{ccc}
\quad\quad\quad\quad FM & \quad\quad\quad\quad FI & \quad\quad\quad\quad FIM \\ 
﻿$\quad \mathbf{M}_{X}=
\begin{bmatrix}   0.0954& 0.2581& 0.6465\\(0.0048)& (0.0077)& (0.0075) \end{bmatrix}$ &
﻿$\quad \mathbf{M}_{X}=
\begin{bmatrix}   0.0910& 0.3862& 0.5228\\(0.0061)& (0.0099)& (0.0097) \end{bmatrix}$ & 
﻿$\quad \mathbf{M}_{X}=
\begin{bmatrix}   0.0893& 0.2711& 0.6396\\(0.0066)& (0.0100)& (0.0116) \end{bmatrix}$  \\& & \\ 
$\mathbf{M}_{X|X^{*}}=
\begin{bmatrix}   0.0795& 0.1049& 0.0907\\(0.0206)& (0.0122)& (0.0093)\\0.7221& 0.2812& 0.0907\\(0.0886)& (0.0293)& (0.0130)\\0.1985& 0.6138& 0.8186\\(0.0801)& (0.0316)& (0.0133) \end{bmatrix}$ &
$\mathbf{M}_{X|X^{*}}=
\begin{bmatrix}   0.1334& 0.0698& 0.0757\\(0.0169)& (0.0170)& (0.0096)\\0.6732& 0.3968& 0.1629\\(0.0393)& (0.0718)& (0.0190)\\0.1935& 0.5334& 0.7614\\(0.0423)& (0.0687)& (0.0203) \end{bmatrix}$ &
$\mathbf{M}_{X|X^{*}}=
\begin{bmatrix}   0.0884& 0.0950& 0.0848\\(0.0237)& (0.0160)& (0.0157)\\0.6053& 0.2472& 0.0822\\(0.0622)& (0.0586)& (0.0274)\\0.3063& 0.6578& 0.8330\\(0.0680)& (0.0558)& (0.0318) \end{bmatrix}$  \\& & \\ 
$\quad\mathbf{M}_{X^{*}}=
\begin{bmatrix}   0.1333& 0.4369& 0.4297\\(0.0415)& (0.0390)& (0.0451) \end{bmatrix}$ &
$\quad\mathbf{M}_{X^{*}}=
\begin{bmatrix}   0.2973& 0.3061& 0.3966\\(0.0518)& (0.0568)& (0.0469) \end{bmatrix}$ &
$\quad\mathbf{M}_{X^{*}}=
\begin{bmatrix}   0.2518& 0.3466& 0.4016\\(0.0622)& (0.0821)& (0.0723) \end{bmatrix}$  \\& & \\ 

\quad\quad\quad\quad FH & \quad\quad\quad\quad FHM & \quad\quad\quad\quad FHI \\ 
﻿$\quad \mathbf{M}_{X}=
\begin{bmatrix}   0.1472& 0.4402& 0.4125\\(0.0078)& (0.0109)& (0.0105) \end{bmatrix}$ &
﻿$\quad \mathbf{M}_{X}=
\begin{bmatrix}   0.1049& 0.3837& 0.5114\\(0.0063)& (0.0101)& (0.0097) \end{bmatrix}$ & 
﻿$\quad \mathbf{M}_{X}=
\begin{bmatrix}   0.1275& 0.4363& 0.4363\\(0.0089)& (0.0126)& (0.0128) \end{bmatrix}$  \\& & \\ 
$\mathbf{M}_{X|X^{*}}=
\begin{bmatrix}   0.2662& 0.0882& 0.0448\\(0.0284)& (0.0187)& (0.0163)\\0.6216& 0.4497& 0.0924\\(0.0258)& (0.0434)& (0.0342)\\0.1121& 0.4621& 0.8628\\(0.0382)& (0.0523)& (0.0384) \end{bmatrix}$ &
$\mathbf{M}_{X|X^{*}}=
\begin{bmatrix}   0.1807& 0.0545& 0.0903\\(0.0228)& (0.0191)& (0.0208)\\0.6616& 0.3563& 0.0788\\(0.0351)& (0.0584)& (0.0341)\\0.1577& 0.5892& 0.8310\\(0.0486)& (0.0505)& (0.0253) \end{bmatrix}$ &
$\mathbf{M}_{X|X^{*}}=
\begin{bmatrix}   0.2343& 0.0209& 0.0625\\(0.0308)& (0.0165)& (0.0213)\\0.6018& 0.4059& 0.1018\\(0.0240)& (0.0533)& (0.0518)\\0.1638& 0.5732& 0.8358\\(0.0302)& (0.0583)& (0.0474) \end{bmatrix}$  \\& & \\ 
$\quad\mathbf{M}_{X^{*}}=
\begin{bmatrix}   0.3831& 0.4061& 0.2108\\(0.0562)& (0.0440)& (0.0421) \end{bmatrix}$ &
$\quad\mathbf{M}_{X^{*}}=
\begin{bmatrix}   0.3258& 0.4144& 0.2598\\(0.0536)& (0.0557)& (0.0453) \end{bmatrix}$ &
$\quad\mathbf{M}_{X^{*}}=
\begin{bmatrix}   0.4609& 0.3421& 0.1970\\(0.0470)& (0.0527)& (0.0457) \end{bmatrix}$  \\& & \\ 

\quad\quad\quad\quad FHIM & \quad\quad\quad\quad D & \quad\quad\quad\quad DM \\ 
﻿$\quad \mathbf{M}_{X}=
\begin{bmatrix}   0.0889& 0.4066& 0.5044\\(0.0095)& (0.0182)& (0.0200) \end{bmatrix}$ &
﻿$\quad \mathbf{M}_{X}=
\begin{bmatrix}   0.1009& 0.4495& 0.4495\\(0.0156)& (0.0261)& (0.0267) \end{bmatrix}$ & 
﻿$\quad \mathbf{M}_{X}=
\begin{bmatrix}   0.0930& 0.3566& 0.5504\\(0.0185)& (0.0331)& (0.0349) \end{bmatrix}$  \\& & \\ 
$\mathbf{M}_{X|X^{*}}=
\begin{bmatrix}   0.1330& 0.0454& 0.0681\\(0.1364)& (0.1907)& (0.0302)\\0.6509& 0.3574& 0.1130\\(0.1044)& (0.1732)& (0.0493)\\0.2161& 0.5972& 0.8190\\(0.0721)& (0.1579)& (0.0437) \end{bmatrix}$ &
$\mathbf{M}_{X|X^{*}}=
\begin{bmatrix}   0.1512& 0.0586& 0.1483\\(0.0729)& (0.1807)& (0.1157)\\0.7297& 0.4394& 0.0001\\(0.0866)& (0.2258)& (0.1376)\\0.1191& 0.5020& 0.8516\\(0.0635)& (0.1777)& (0.1196) \end{bmatrix}$ &
$\mathbf{M}_{X|X^{*}}=
\begin{bmatrix}   0.0000& 0.8037& 0.0885\\(0.1649)& (0.3519)& (0.0518)\\0.8040& 0.0002& 0.0989\\(0.1377)& (0.2989)& (0.0710)\\0.1960& 0.1961& 0.8127\\(0.0813)& (0.1651)& (0.0681) \end{bmatrix}$  \\& & \\ 
$\quad\mathbf{M}_{X^{*}}=
\begin{bmatrix}   0.4190& 0.2792& 0.3018\\(0.1090)& (0.0932)& (0.0886) \end{bmatrix}$ &
$\quad\mathbf{M}_{X^{*}}=
\begin{bmatrix}   0.2924& 0.5375& 0.1701\\(0.0758)& (0.1799)& (0.1765) \end{bmatrix}$ &
$\quad\mathbf{M}_{X^{*}}=
\begin{bmatrix}   0.3742& 0.0511& 0.5747\\(0.1017)& (0.1431)& (0.0882) \end{bmatrix}$  \\& & \\ 

\end{tabular}
}
\caption{Distribution of reported and latent LS for different socioeconomic
groups.}
\label{table:74}
\end{table}

\begin{table}[H]
\resizebox{\textwidth}{!}{\begin{tabular}{ccc}
\quad\quad\quad\quad DI & \quad\quad\quad\quad DIM & \quad\quad\quad\quad DH \\ 
﻿$\quad \mathbf{M}_{X}=
\begin{bmatrix}   0.0515& 0.3557& 0.5928\\(0.0092)& (0.0176)& (0.0183) \end{bmatrix}$ &
﻿$\quad \mathbf{M}_{X}=
\begin{bmatrix}   0.0545& 0.2497& 0.6958\\(0.0059)& (0.0099)& (0.0110) \end{bmatrix}$ & 
﻿$\quad \mathbf{M}_{X}=
\begin{bmatrix}   0.1667& 0.4825& 0.3509\\(0.0361)& (0.0480)& (0.0468) \end{bmatrix}$  \\& & \\ 
$\mathbf{M}_{X|X^{*}}=
\begin{bmatrix}   0.0004& 0.7156& 0.0490\\(0.1492)& (0.3080)& (0.0164)\\0.7196& 0.0042& 0.1623\\(0.1090)& (0.2801)& (0.0217)\\0.2800& 0.2802& 0.7888\\(0.1075)& (0.1254)& (0.0281) \end{bmatrix}$ &
$\mathbf{M}_{X|X^{*}}=
\begin{bmatrix}   0.0646& 0.0170& 0.1040\\(0.0423)& (0.0195)& (0.0284)\\0.5868& 0.2153& 0.0007\\(0.1018)& (0.0977)& (0.0331)\\0.3486& 0.7677& 0.8953\\(0.1127)& (0.1066)& (0.0309) \end{bmatrix}$ &
$\mathbf{M}_{X|X^{*}}=
\begin{bmatrix}   0.2964& 0.1344& 0.0000\\(0.2072)& (0.2804)& (0.0847)\\0.6500& 0.2666& 0.4009\\(0.2099)& (0.2688)& (0.1502)\\0.0536& 0.5990& 0.5990\\(0.0451)& (0.1835)& (0.1392) \end{bmatrix}$  \\& & \\ 
$\quad\mathbf{M}_{X^{*}}=
\begin{bmatrix}   0.3563& 0.0289& 0.6148\\(0.1148)& (0.1315)& (0.0589) \end{bmatrix}$ &
$\quad\mathbf{M}_{X^{*}}=
\begin{bmatrix}   0.2598& 0.4506& 0.2896\\(0.0526)& (0.0852)& (0.1172) \end{bmatrix}$ &
$\quad\mathbf{M}_{X^{*}}=
\begin{bmatrix}   0.4550& 0.2597& 0.2853\\(0.1136)& (0.1337)& (0.1251) \end{bmatrix}$  \\& & \\ 

\quad\quad\quad\quad DHM & \quad\quad\quad\quad DHI & \quad\quad\quad\quad DHIM \\ 
﻿$\quad \mathbf{M}_{X}=
\begin{bmatrix}   0.0952& 0.4218& 0.4830\\(0.0245)& (0.0389)& (0.0417) \end{bmatrix}$ &
﻿$\quad \mathbf{M}_{X}=
\begin{bmatrix}   0.0764& 0.4509& 0.4727\\(0.0174)& (0.0287)& (0.0275) \end{bmatrix}$ & 
﻿$\quad \mathbf{M}_{X}=
\begin{bmatrix}   0.0506& 0.3244& 0.6250\\(0.0100)& (0.0199)& (0.0189) \end{bmatrix}$  \\& & \\ 
$\mathbf{M}_{X|X^{*}}=
\begin{bmatrix}   0.8068& 0.0000& 0.0000\\(0.3230)& (0.2582)& (0.0000)\\0.1931& 0.7450& 0.1388\\(0.3289)& (0.3020)& (0.0881)\\0.0001& 0.2550& 0.8612\\(0.0740)& (0.1666)& (0.0881) \end{bmatrix}$ &
$\mathbf{M}_{X|X^{*}}=
\begin{bmatrix}   0.1842& 0.0000& 0.0586\\(0.0609)& (0.1091)& (0.0400)\\0.7490& 0.4135& 0.1931\\(0.0664)& (0.1295)& (0.0883)\\0.0667& 0.5864& 0.7482\\(0.0428)& (0.1241)& (0.0759) \end{bmatrix}$ &
$\mathbf{M}_{X|X^{*}}=
\begin{bmatrix}   0.1110& 0.0463& 0.0305\\(0.0531)& (0.0361)& (0.0194)\\0.7734& 0.3353& 0.1030\\(0.0818)& (0.0839)& (0.0433)\\0.1156& 0.6184& 0.8666\\(0.0814)& (0.0645)& (0.0408) \end{bmatrix}$  \\& & \\ 
$\quad\mathbf{M}_{X^{*}}=
\begin{bmatrix}   0.1180& 0.4563& 0.4257\\(0.1225)& (0.1467)& (0.0990) \end{bmatrix}$ &
$\quad\mathbf{M}_{X^{*}}=
\begin{bmatrix}   0.3155& 0.3737& 0.3108\\(0.0473)& (0.1092)& (0.1135) \end{bmatrix}$ &
$\quad\mathbf{M}_{X^{*}}=
\begin{bmatrix}   0.1451& 0.5341& 0.3207\\(0.0471)& (0.1166)& (0.0989) \end{bmatrix}$  \\& & \\ 

\quad\quad\quad\quad DF & \quad\quad\quad\quad DFM & \quad\quad\quad\quad DFI \\ 
﻿$\quad \mathbf{M}_{X}=
\begin{bmatrix}   0.0628& 0.3744& 0.5628\\(0.0120)& (0.0218)& (0.0227) \end{bmatrix}$ &
﻿$\quad \mathbf{M}_{X}=
\begin{bmatrix}   0.0549& 0.2517& 0.6934\\(0.0086)& (0.0166)& (0.0184) \end{bmatrix}$ & 
﻿$\quad \mathbf{M}_{X}=
\begin{bmatrix}   0.0631& 0.3157& 0.6211\\(0.0062)& (0.0136)& (0.0136) \end{bmatrix}$  \\& & \\ 
$\mathbf{M}_{X|X^{*}}=
\begin{bmatrix}   0.3225& 0.0341& 0.0199\\(0.2144)& (0.3108)& (0.0192)\\0.6770& 0.5689& 0.1129\\(0.1648)& (0.2530)& (0.0439)\\0.0005& 0.3970& 0.8672\\(0.1045)& (0.1589)& (0.0481) \end{bmatrix}$ &
$\mathbf{M}_{X|X^{*}}=
\begin{bmatrix}   0.0871& 0.0002& 0.0680\\(0.1509)& (0.2390)& (0.0329)\\0.6268& 0.3025& 0.0860\\(0.1473)& (0.1473)& (0.0446)\\0.2861& 0.6973& 0.8459\\(0.1290)& (0.1737)& (0.0534) \end{bmatrix}$ &
$\mathbf{M}_{X|X^{*}}=
\begin{bmatrix}   0.1041& 0.0538& 0.0495\\(0.0375)& (0.0322)& (0.0124)\\0.6333& 0.3570& 0.0517\\(0.1162)& (0.1273)& (0.0315)\\0.2626& 0.5893& 0.8988\\(0.1095)& (0.1206)& (0.0336) \end{bmatrix}$  \\& & \\ 
$\quad\mathbf{M}_{X^{*}}=
\begin{bmatrix}   0.1218& 0.4228& 0.4554\\(0.1327)& (0.1535)& (0.0783) \end{bmatrix}$ &
$\quad\mathbf{M}_{X^{*}}=
\begin{bmatrix}   0.2057& 0.2514& 0.5429\\(0.0773)& (0.1476)& (0.1384) \end{bmatrix}$ &
$\quad\mathbf{M}_{X^{*}}=
\begin{bmatrix}   0.2142& 0.4569& 0.3289\\(0.0649)& (0.1051)& (0.0854) \end{bmatrix}$  \\& & \\ 
\end{tabular}
}
\caption{Distribution of reported and latent LS for different socioeconomic
groups.}
\label{table:75}
\end{table}

\begin{table}[H]
\resizebox{\textwidth}{!}{\begin{tabular}{ccc}
\quad\quad\quad\quad DFIM & \quad\quad\quad\quad DFH & \quad\quad\quad\quad DFHM \\ 
﻿$\quad \mathbf{M}_{X}=
\begin{bmatrix}   0.0647& 0.2184& 0.7169\\(0.0077)& (0.0119)& (0.0130) \end{bmatrix}$ &
﻿$\quad \mathbf{M}_{X}=
\begin{bmatrix}   0.1531& 0.4439& 0.4031\\(0.0258)& (0.0375)& (0.0347) \end{bmatrix}$ & 
﻿$\quad \mathbf{M}_{X}=
\begin{bmatrix}   0.0495& 0.3357& 0.6148\\(0.0113)& (0.0293)& (0.0294) \end{bmatrix}$  \\& & \\ 
$\mathbf{M}_{X|X^{*}}=
\begin{bmatrix}   0.0716& 0.0608& 0.0643\\(0.0800)& (0.0359)& (0.0174)\\0.5391& 0.3101& 0.0642\\(0.1563)& (0.1190)& (0.0271)\\0.3893& 0.6291& 0.8715\\(0.1624)& (0.1243)& (0.0397) \end{bmatrix}$ &
$\mathbf{M}_{X|X^{*}}=
\begin{bmatrix}   0.2893& 0.0509& 0.0000\\(0.0883)& (0.2123)& (0.0060)\\0.6115& 0.4170& 0.0558\\(0.0967)& (0.1764)& (0.0827)\\0.0992& 0.5321& 0.9442\\(0.0502)& (0.1811)& (0.0852) \end{bmatrix}$ &
$\mathbf{M}_{X|X^{*}}=
\begin{bmatrix}   0.0006& 0.6587& 0.0203\\(0.1922)& (0.2721)& (0.0193)\\0.6584& 0.0001& 0.1001\\(0.1871)& (0.2395)& (0.0525)\\0.3410& 0.3411& 0.8796\\(0.1478)& (0.2120)& (0.0515) \end{bmatrix}$  \\& & \\ 
$\quad\mathbf{M}_{X^{*}}=
\begin{bmatrix}   0.1849& 0.2699& 0.5452\\(0.0690)& (0.1118)& (0.1230) \end{bmatrix}$ &
$\quad\mathbf{M}_{X^{*}}=
\begin{bmatrix}   0.4662& 0.3572& 0.1766\\(0.0729)& (0.0981)& (0.1010) \end{bmatrix}$ &
$\quad\mathbf{M}_{X^{*}}=
\begin{bmatrix}   0.4335& 0.0582& 0.5084\\(0.1373)& (0.1495)& (0.0994) \end{bmatrix}$  \\& & \\ 

\quad\quad\quad\quad DFHI & \quad\quad\quad\quad DFHIM & \quad\quad\quad\quad  \\ 
﻿$\quad \mathbf{M}_{X}=
\begin{bmatrix}   0.0825& 0.4599& 0.4575\\(0.0118)& (0.0219)& (0.0231) \end{bmatrix}$ &
﻿$\quad \mathbf{M}_{X}=
\begin{bmatrix}   0.0564& 0.3725& 0.5711\\(0.0123)& (0.0234)& (0.0241) \end{bmatrix}$ & 
\\ & & \\ 
$\mathbf{M}_{X|X^{*}}=
\begin{bmatrix}   0.1694& 0.0132& 0.1297\\(0.0392)& (0.0268)& (0.0512)\\0.7000& 0.4105& 0.0642\\(0.0487)& (0.0558)& (0.0648)\\0.1306& 0.5763& 0.8060\\(0.0569)& (0.0498)& (0.0643) \end{bmatrix}$ &
$\mathbf{M}_{X|X^{*}}=
\begin{bmatrix}   0.0855& 0.0354& 0.0485\\(0.0418)& (0.1193)& (0.0236)\\0.6881& 0.3458& 0.0945\\(0.1017)& (0.1544)& (0.0470)\\0.2264& 0.6187& 0.8570\\(0.1027)& (0.1612)& (0.0579) \end{bmatrix}$ &
\\ & & \\ 
$\quad\mathbf{M}_{X^{*}}=
\begin{bmatrix}   0.3391& 0.5201& 0.1408\\(0.0518)& (0.0749)& (0.0508) \end{bmatrix}$ &
$\quad\mathbf{M}_{X^{*}}=
\begin{bmatrix}   0.3298& 0.3272& 0.3431\\(0.0819)& (0.0954)& (0.0605) \end{bmatrix}$ &
\\& & \\ 
\end{tabular}
}
\caption{Distribution of reported and latent LS for different socioeconomic
groups.}
\label{table:76}
\end{table}
\newpage 
\bibliographystyle{apalike}
\bibliography{Review}

\begin{thebibliography}{}

\bibitem[Andrews, 1999]{andrews_estimation_1999}
Andrews, D. W.~K. (1999).
\newblock Estimation {When} a {Parameter} is on a {Boundary}.
\newblock {\em Econometrica}, 67(6):1341--1383.

\bibitem[Andrews, 2000]{andrews_inconsistency_2000}
Andrews, D. W.~K. (2000).
\newblock Inconsistency of the {Bootstrap} {When} a {Parameter} is on the
  {Boundary} of the {Parameter} {Space}.
\newblock {\em Econometrica}, 68(2):399--405.

\bibitem[Andrews, 2001]{andrews_testing_2001}
Andrews, D. W.~K. (2001).
\newblock Testing {When} a {Parameter} {Is} on the {Boundary} of the
  {Maintained} {Hypothesis}.
\newblock {\em Econometrica}, 69(3):683--734.

\bibitem[Barrington-Leigh and Behzadnejad, 2017]{barrington-leigh_impact_2017}
Barrington-Leigh, C. and Behzadnejad, F. (2017).
\newblock The impact of daily weather conditions on life satisfaction:
  {Evidence} from cross-sectional and panel data.
\newblock {\em Journal of Economic Psychology}, 59:145--163.

\bibitem[Bertrand and Mullainathan, 2001]{bertrand_people_2001}
Bertrand, M. and Mullainathan, S. (2001).
\newblock Do {People} {Mean} {What} {They} {Say}? {Implications} for
  {Subjective} {Survey} {Data}.
\newblock {\em American Economic Review}, 91(2):67--72.

\bibitem[Bond and Lang, 2014]{bond_sad_2014}
Bond, T. and Lang, K. (2014).
\newblock The {Sad} {Truth} {About} {Happiness} {Scales}.
\newblock Technical Report w19950, National Bureau of Economic Research,
  Cambridge, MA.

\bibitem[Bond and Lang, 2018]{bond_sad_nodate}
Bond, T. and Lang, K. (2018).
\newblock The {Sad} {Truth} {About} {Happiness} {Scales}.
\newblock {\em Journal of Political Economy (forthcoming)}.

\bibitem[Bound and Krueger, 1991]{bound_extent_1991}
Bound, J. and Krueger, A.~B. (1991).
\newblock The {Extent} of {Measurement} {Error} in {Longitudinal} {Earnings}
  {Data}: {Do} {Two} {Wrongs} {Make} a {Right}?
\newblock {\em Journal of Labor Economics}, 9(1):1--24.

\bibitem[Chen et~al., 2019]{chen_have_2019}
Chen, L.~Y., Oparina, E., Powdthavee, N., and Srisuma, S. (2019).
\newblock Have {Econometric} {Analyses} of {Happiness} {Data} {Been} {Futile}?
  {A} {Simple} {Truth} {About} {Happiness} {Scales}.
\newblock {\em Working paper}.

\bibitem[Chen and Khan, 2003]{chen_rates_2003}
Chen, S. and Khan, S. (2003).
\newblock Rates of convergence for estimating regression coefficients in
  heteroskedastic discrete response models.
\newblock {\em Journal of Econometrics}, 117(2):245--278.

\bibitem[Chen et~al., 2005]{chen_measurement_2005}
Chen, X., Hong, H., and Tamer, E. (2005).
\newblock Measurement {Error} {Models} with {Auxiliary} {Data}.
\newblock {\em The Review of Economic Studies}, 72(2):343--366.

\bibitem[Clark et~al., 2008]{clark_relative_2008}
Clark, A.~E., Frijters, P., and Shields, M.~A. (2008).
\newblock Relative {Income}, {Happiness}, and {Utility}: {An} {Explanation} for
  the {Easterlin} {Paradox} and {Other} {Puzzles}.
\newblock {\em Journal of Economic Literature}, 46(1):95--144.

\bibitem[Di~Tella et~al., 2001]{di_tella_preferences_2001}
Di~Tella, R., MacCulloch, R.~J., and Oswald, A.~J. (2001).
\newblock Preferences over {Inflation} and {Unemployment}: {Evidence} from
  {Surveys} of {Happiness}.
\newblock {\em The American Economic Review}, 91(1):335--341.

\bibitem[Diener, 2009]{diener_assessing_2009}
Diener, E. (2009).
\newblock {\em Assessing {Well}-{Being}: {The} {Collected} {Works} of {Ed}
  {Diener}}.
\newblock Springer Science \& Business Media.

\bibitem[Diener et~al., 1985]{diener_satisfaction_1985}
Diener, E., Emmons, R.~A., Larsen, R.~J., and Griffin, S. (1985).
\newblock The {Satisfaction} {With} {Life} {Scale}.
\newblock {\em Journal of Personality Assessment}, 49(1):71--75.

\bibitem[Diener et~al., 2009]{diener_subjective_2009}
Diener, E., Oishi, S., and Lucas, R. (2009).
\newblock Subjective {Well}-{Being}: {The} {Science} of {Happiness} and {Life}
  {Satisfaction}.
\newblock {\em The Oxford Handbook of Positive Psychology}.

\bibitem[Diener et~al., 2018]{diener_advances_2018}
Diener, E., Oishi, S., and Tay, L. (2018).
\newblock Advances in subjective well-being research.
\newblock {\em Nature Human Behaviour}.

\bibitem[Diener et~al., 1991]{diener_response_1991}
Diener, E., Sandvik, E., Pavot, W., and Gallagher, D. (1991).
\newblock Response {Artifacts} in the {Measurement} of {Subjective}
  {Well}-{Being}.
\newblock {\em Social Indicators Research}, 24(1):35--56.

\bibitem[Diener et~al., 2017]{diener_findings_2017}
Diener, E., Tay, L., Heintzelman, S.~J., Kushlev, K., Wirtz, D., and Lutes,
  L.~D. (2017).
\newblock Findings {All} {Psychologists} {Should} {Know} {From} the {New}
  {Science} on {Subjective} {Well}-{Being}.
\newblock {\em Canadian Psychology/Psychologie canadienne}, 58(2):87--104.

\bibitem[Dolan et~al., 2008]{dolan_we_2008}
Dolan, P., Peasgood, T., and White, M. (2008).
\newblock Do we really know what makes us happy? {A} review of the economic
  literature on the factors associated with subjective well-being.
\newblock {\em Journal of Economic Psychology}, 29(1):94--122.

\bibitem[Easterlin, 1974]{easterlin_does_1974}
Easterlin, R.~A. (1974).
\newblock Does economic growth improve the human lot? : some empirical
  evidence.
\newblock {\em Nations and households in economic growth : essays in honor of
  Moses Abramovitz}.

\bibitem[Easterlin, 1995]{easterlin_will_1995}
Easterlin, R.~A. (1995).
\newblock Will raising the incomes of all increase the happiness of all?
\newblock {\em Journal of Economic Behavior \& Organization}, 27(1):35--47.

\bibitem[Feddersen et~al., 2016]{feddersen_subjective_2016}
Feddersen, J., Metcalfe, R., and Wooden, M. (2016).
\newblock Subjective wellbeing: why weather matters.
\newblock {\em Journal of the Royal Statistical Society: Series A (Statistics
  in Society)}, 179(1):203--228.

\bibitem[{Ferrer-i-Carbonell} and Frijters, 2004]{ferrericarbonell_how_2004}
{Ferrer-i-Carbonell}, A. and Frijters, P. (2004).
\newblock How {Important} is {Methodology} for the estimates of the
  determinants of {Happiness}?
\newblock {\em The Economic Journal}, 114(497):641--659.

\bibitem[Gin\`{e} and Zinn, 1990]{gin`e_bootstrapping_1990}
Gin\`{e}, E. and Zinn, J. (1990).
\newblock Bootstrapping {General} {Empirical} {Measures}.
\newblock {\em The Annals of Probability}, 18(2):851--869.

\bibitem[Gruber and Mullainathan, 2006]{gruber_cigarette_2006}
Gruber, J. and Mullainathan, S. (2006).
\newblock Do {Cigarette} {Taxes} {Make} {Smokers} {Happier}?
\newblock In Ng, Y.-K. and Ho, L.~S., editors, {\em Happiness and {Public}
  {Policy}: {Theory}, {Case} {Studies} and {Implications}}, pages 109--146.
  Palgrave Macmillan UK, London.

\bibitem[Heffetz and Rabin, 2013]{heffetz_conclusions_2013}
Heffetz, O. and Rabin, M. (2013).
\newblock Conclusions {Regarding} {Cross}-{Group} {Differences} in {Happiness}
  {Depend} on {Difficulty} of {Reaching} {Respondents}.
\newblock {\em American Economic Review}, 103(7):3001--3021.

\bibitem[Helliwell et~al., 2012]{helliwell_world_2012}
Helliwell, J., Layard, R., and Sachs, J. (2012).
\newblock World happiness report.
\newblock {LSE} {Research} {Online} {Documents} on {Economics}, London School
  of Economics and Political Science, LSE Library.

\bibitem[Hu, 2008]{hu_identification_2008}
Hu, Y. (2008).
\newblock Identification and estimation of nonlinear models with
  misclassification error using instrumental variables: {A} general solution.
\newblock {\em Journal of Econometrics}, 144(1):27--61.

\bibitem[Hu, 2017]{hu_econometrics_2017}
Hu, Y. (2017).
\newblock The econometrics of unobservables: {Applications} of measurement
  error models in empirical industrial organization and labor economics.
\newblock {\em Journal of Econometrics}, 200(2):154--168.

\bibitem[Hu and Schennach, 2008]{hu_instrumental_2008}
Hu, Y. and Schennach, S.~M. (2008).
\newblock Instrumental {Variable} {Treatment} of {Nonclassical} {Measurement}
  {Error} {Models}.
\newblock {\em Econometrica}, 76(1):195--216.

\bibitem[Kahneman et~al., 1999]{kahneman_well-being:_1999}
Kahneman, D., Diener, E., and Schwarz, N., editors (1999).
\newblock {\em Well-being: {The} foundations of hedonic psychology.}
\newblock Well-being: {The} foundations of hedonic psychology. Russell Sage
  Foundation, New York, NY, US.

\bibitem[Lee, 1992]{lee_median_1992}
Lee, M.~J. (1992).
\newblock Median regression for ordered discrete response.
\newblock {\em Journal of Econometrics}, 51(1):59--77.

\bibitem[Mahajan, 2006]{mahajan_identification_2006}
Mahajan, A. (2006).
\newblock Identification and {Estimation} of {Regression} {Models} with
  {Misclassification}.
\newblock {\em Econometrica}, 74(3):631--665.

\bibitem[Manski, 1985]{manski_semiparametric_1985}
Manski, C. (1985).
\newblock Semiparametric analysis of discrete response: {Asymptotic} properties
  of the maximum score estimator.
\newblock {\em Journal of Econometrics}, 27(3):313--333.

\bibitem[McCrae and Costa, 1987]{mccrae_validation_1987}
McCrae, R.~R. and Costa, P.~T. (1987).
\newblock Validation of the five-factor model of personality across instruments
  and observers.
\newblock {\em Journal of Personality and Social Psychology}, 52(1):81--90.

\bibitem[Meisenberg and Woodley, 2015]{meisenberg_gender_2015}
Meisenberg, G. and Woodley, M.~A. (2015).
\newblock Gender {Differences} in {Subjective} {Well}-{Being} and {Their}
  {Relationships} with {Gender} {Equality}.
\newblock {\em Journal of Happiness Studies}, 16(6):1539--1555.

\bibitem[{OECD}, 2013]{oecd_oecd_2013}
{OECD} (2013).
\newblock {\em {OECD} {Guidelines} on {Measuring} {Subjective} {Well}-being}.

\bibitem[Oswald and Powdthavee, 2008]{oswald_does_2008}
Oswald, A.~J. and Powdthavee, N. (2008).
\newblock Does happiness adapt? {A} longitudinal study of disability with
  implications for economists and judges.
\newblock {\em Journal of Public Economics}, 92(5):1061--1077.

\bibitem[Schennach, 2013]{schennach_2013}
Schennach, S.~M. (2013).
\newblock {\em Measurement Error in Nonlinear Models – A Review}, volume~3 of
  {\em Econometric Society Monographs}, pages 296--–337.
\newblock Cambridge University Press.

\bibitem[Schr\"{o}der and Yitzhaki, 2017]{schroder_revisiting_2017}
Schr\"{o}der, C. and Yitzhaki, S. (2017).
\newblock Revisiting the evidence for cardinal treatment of ordinal variables.
\newblock {\em European Economic Review}, 92:337--358.

\bibitem[Schwarz, 2014]{schwarz_cognition_2014}
Schwarz, N. (2014).
\newblock {\em Cognition and {Communication}: {Judgmental} {Biases}, {Research}
  {Methods}, and the {Logic} of {Conversation}}.
\newblock Psychology Press.

\bibitem[Schwarz and Clore, 1983]{schwarz_mood_1983}
Schwarz, N. and Clore, G.~L. (1983).
\newblock Mood, misattribution, and judgments of well-being: {Informative} and
  directive functions of affective states.
\newblock {\em Journal of Personality and Social Psychology}, 45(3):513--523.

\bibitem[Stevenson and Wolfers, 2009]{stevenson_paradox_2009}
Stevenson, B. and Wolfers, J. (2009).
\newblock The {Paradox} of {Declining} {Female} {Happiness}.
\newblock {\em American Economic Journal: Economic Policy}, 1(2):190--225.

\bibitem[Strack et~al., 1991]{strack_subjective_1991}
Strack, F., Argyle, M., and Schwarz, N. (1991).
\newblock {\em Subjective {Well}-being: {An} {Interdisciplinary}
  {Perspective}}.
\newblock International series in experimental social psychology. Pergamon
  Press.

\bibitem[Van~Praag, 1971]{van_praag_welfare_1971}
Van~Praag, B. (1971).
\newblock The welfare function of income in {Belgium}: {An} empirical
  investigation.
\newblock {\em European Economic Review}, 2(3):337--369.

\bibitem[Wilhelm, 2018]{wilhelm_testing_2018}
Wilhelm, D. (2018).
\newblock Testing for the {Presence} of {Measurement} {Error}.
\newblock {\em CeMMAP Working Paper CWP45/18}.

\bibitem[Williams, 2010]{williams_fitting_2010}
Williams, R. (2010).
\newblock Fitting heterogeneous choice models with oglm.
\newblock {\em The Stata Journal}, 10(4):540--567.

\bibitem[Winkelmann and Winkelmann, 1998]{winkelmann_why_1998}
Winkelmann, L. and Winkelmann, R. (1998).
\newblock Why are the {Unemployed} {So} {Unhappy}? {Evidence} from {Panel}
  {Data}.
\newblock {\em Economica}, 65(257):1--15.

\end{thebibliography}

\end{document}